\documentclass[a4paper,11pt]{article}

\usepackage{jheppub} 

\usepackage[T1]{fontenc} 

\usepackage{pstricks}
\usepackage{enumerate}
\usepackage{color}
\usepackage{slashed}
\usepackage{graphicx,graphics}
\usepackage{verbatim}
\usepackage{amssymb}
\usepackage{amsthm}
\usepackage{array}
\usepackage{amsmath,amsfonts}
\usepackage{caption}
\usepackage[caption=false]{subfig}
\usepackage{multirow}
\usepackage{dcolumn}
\usepackage{bm}
\usepackage{color}
 \usepackage{epstopdf}
\long\def\rpl#1!!#2!!{\textcolor{red}{#1} \textcolor{blue}{#2}}

\def\m00{\mu_{e e}}

\def\m0p{\mu_{\mu e}}

\def\m0m{\mu_{e 	au}}
\def\O11{O_{\mu 1}}
\def\r12{O_{\mu 2}}
\def\r13{O_{\mu 3}}
\def\r21{O_{e 1}}
\def\r22{O_{e 2}}
\def\r23{O_{e 3}}
\def\r31{O_{	tau 1}}
\def\r32{O_{	tau 2}}
\def\r33{O_{	tau 3}}
\def\a{\alpha}

\def\b{\beta}

\def\l{\lambda}
\def\m{\mu}

\def\q2 {q^2}

\def\r {\rightarrow}

\def\bt{\begin{table}}
\def\et{\end{table}}

\newcommand{\ba}{\begin{array}}
\newcommand{\ea}{\end{array}}
\newcommand{\bd}{\begin{displaymath}}
\newcommand{\ed}{\end{displaymath}}
\newcommand{\besub}{\begin{subequations}}
\newcommand{\eesub}{\end{subequations}}
\newcommand{\be}{\begin{equation}}
\newcommand{\ee}{\end{equation}}
\newcommand{\bea}{\begin{eqnarray}}
\newcommand{\eea}{\end{eqnarray}}

\allowdisplaybreaks

\title{\boldmath Charged scalars confronting neutrino mass and muon $g-2$ anomaly}

\author[a]{Nabarun Chakrabarty,}
\author[b,a,c]{~Cheng-Wei Chiang,}
\author[d]{~Takahiro Ohata,}
\author[d]{~Koji Tsumura}

\affiliation[a]{Physics Division, National Center for Theoretical Sciences, Hsinchu, Taiwan 30013, R.O.C.}
\affiliation[b]{Department of Physics, National Taiwan University,
Taipei, Taiwan 10617, R.O.C.}
\affiliation[c]{Institute of Physics, Academia Sinica,
		Taipei, Taiwan 11529, R.O.C.}
\affiliation[d]{Department of Physics, Kyoto University,
Kyoto 606-8502, Japan}

\emailAdd{nchakrabarty@cts.nthu.edu.tw}
\emailAdd{chengwei@phys.ntu.edu.tw}
\emailAdd{tk.ohata@gauge.scphys.kyoto-u.ac.jp}
\emailAdd{ko2@gauge.scphys.kyoto-u.ac.jp}

\abstract{The present work introduces two possible extensions of the Standard Model Higgs sector. 
In the first case, the Zee-Babu type model for the generation of neutrino mass is augmented with a scalar triplet and additional singly charged scalar singlets. The second scenario, on the other hand, generalizes the Type-II seesaw model by replicating the number of the scalar triplets. A $\mathbb{Z}_3$ symmetry is imposed in case of both the scenarios, but, allowed to be violated by terms of mass dimension two and three for generating neutrino masses and mixings. We examine how the models so introduced can explain the experimental observation on the muon anomalous magnetic moment. We estimate the two-loop contribution to neutrino mass induced by the scalar triplet, in addition to what comes from the doubly charged singlet in the usual Zee-Babu framework, in the first model. On the other hand, the neutrino mass arises in the usual Type-II fashion in the second model. In addition, the role of the $\mathbb{Z}_3$ symmetry in suppressing lepton flavor violation is also elucidated. }

\preprint{NCTS-PH/1811,~KUNS-2733}

\begin{document} 
\maketitle
\flushbottom

\section{Introduction}\label{intro}

Evidence of a sizeable deviation in the measured muon anomalous magnetic moment from its Standard Model (SM) expectation is likely to call for physics beyond the SM.  A 3.6$\,\sigma$ discrepancy between theoretical calculations within the SM and 
experimental data~\cite{Bennett:2006fi}, quoting
\bea
\Delta a_{\mu} = a_\mu^{\text{exp}} - a_\mu^{\text{SM}} = 288(63)(48) \times 10^{-11}
~.
\eea
Another important issue is the inability to generate non-zero neutrino mass within the SM.  While non-zero neutrino mass can be induced at tree level using the Type-I~\cite{Minkowski:1977sc,Sawada:1979dis,GellMann:1980vs,Glashow:1979nm,Mohapatra:1979ia}, Type-II~\cite{PhysRevD.22.2227,Magg:1980ut,Lazarides:1980nt} and Type-III~\cite{Foot:1988aq} seesaw mechanisms, an also attractive way in this context is to invoke loop processes~\cite{Zee:1980ai,Babu:1988ki,Zee:1985rj,Zee:1985id} for the same.  (See also Refs.~\cite{Farzan:2012ev,Angel:2012ug,Law:2013dya,Cai:2017jrq,Sugiyama:2015cra,ANTIPIN2017330} for recent reviews.)  In such a case, the scale of the new physics responsible for generating neutrino mass can be not too far from the TeV scale, thereby enhancing the observability at colliders.  We pick up two such scenarios that are particularly relevant for the present discussion.  These are the Type-II seesaw scenario that employs a scalar $SU(2)_L$ triplet~\cite{PhysRevD.22.2227,LAZARIDES1981287,PhysRevD.23.165} and the Zee-Babu model~\cite{Zee:1980ai,Babu:1988ki} that introduces two $SU(2)_L$ singlet scalars that carry one and two units of electric charge, respectively.  However, the Type-II seesaw model has been ruled out due to a negative contribution to the muon anomalous magnetic moment~\cite{Fukuyama:2009xk}. The Zee-Babu model also does not fare well in this direction owing to the constraints put on it by non-observation of various lepton flavor-violating decays~\cite{Schmidt:2014zoa,Herrero-Garcia:2014hfa}.

In this paper, we propose two models that serve as unified frameworks to address the muon $g-2$ anomaly and the current data on neutrino masses and mixings. A common feature of two models is the simultaneous existence of doubly charged scalars in 
the $SU(2)_{L}$ singlet and triplet. 
Thanks to their right-chiral and left-chiral Yukawa interactions 
and also non-zero mixing between the doubly charged scalar states, 
the experimentally favored sign of the anomalous muon $g-2$ deviation is achieved. 
Furthermore, the overall magnitude of the contribution is enhanced by the chirality flipping effect. 
On the other hand, the presence of two doubly charged scalars suffer severe constraints 
from the non-observation of the lepton flavor violating processes. 
We will show that the lepton flavor-violating decays turn out to be naturally suppressed in these models 
by imposing a  (softly broken) global $\mathbb{Z}_3$ symmetry without spoiling the explanation of the muon $g-2$ anomaly.

 In addition to the SM fields, the first model features a scalar $SU(2)_L$ triplet, a doubly charged scalar singlet and three singly charged scalar singlets.  The singly charged scalars are charged under the global $\mathbb{Z}_3$
while the triplet and the doubly charged singlet remain neutral.  The Weinberg operator~\cite{Weinberg:1979sa} responsible for neutrino mass can be derived in this model at the two-loop level, similar to what happens in the Zee-Babu model
\footnote{Refs.~\cite{Schmidt:2014zoa,Herrero-Garcia:2014hfa,Chao:2012xt,OHLSSON2009269} are recent studies on the Zee-Babu model.  Some variants of the original model can be seen in Refs.~\cite{Baek:2012ub,Guo:2017gxp,Nomura:2016ask,Okada:2014qsa}.}. 
The second model features three $SU(2)_L$ triplet scalars that are distinguished from one another by their $\mathbb{Z}_3$ charges.  In addition, a $\mathbb{Z}_3$-neutral doubly charged $SU(2)_L$ singlet scalar is also present.  A small neutrino mass arises in this model when the scalar triplets acquire vacuum expectation values (VEV's) to mimic the usual Type-II seesaw model. 
%
%
Besides, in the case of the first model, the same mixing also induces sizeable contributions to the neutrino mass elements through two-loop amplitudes. 
Therefore, the proposed models emerge as novel scenarios successfully connecting the observation of small but non-zero neutrino mass with the long-standing muon $g-2$ anomaly, without invoking additional fermionic degrees of freedom. Further, we note in passing that it is possible to identify appropriate collider signatures that can potentially distinguish the models discussed here from the usual Type-II seesaw model.

This paper is organized as follows. In Sections~\ref{model_A} and \ref{model_B}, we introduce the two models, discussing the additional scalar content in them and the assignment of the global symmetry charges. For the first model, we discuss the contribution of the given scenario to the muon anomalous magnetic moment in Section~\ref{muon_g-2} and explain the current discrepancy between experimental data and the SM expectation. Appropriate discussions on various lepton flavor-violating decays can be found in the same section. Section~\ref{numass} outlines the calculation of neutrino mass, and identifies the parameter space allowed by the recent neutrino data. The numerical results for the second model are 
detailed in Section~\ref{results_B}.
The results obtained are summarized in Section~\ref{summary}. Important expressions encountered while calculating the two-loop neutrino mass matrix are relegated to the Appendix.

\section{Models
}\label{model}

\subsection{Scenario A: Two-loop realization
}\label{model_A}

In this model, the scalar sector of the SM is augmented by an $SU(2)_L$ scalar triplet $\Delta$, a doubly charged scalar singlet $k^{++}$ and three singly charged scalar singlets $k^+_{\mu},k^+_e,k^+_\tau$.\footnote{A recent study also with singly charged scalars in the Zee-Babu context is Ref.~\cite{Nomura:2018vfz}} 
A $\mathbb{Z}_3$ symmetry is imposed, whose utility will become clear in the subsequent sections. Tables~\ref{caseA_sm} and \ref{caseA_bsm} list the quantum numbers of both SM and additional fields respectively.

\begin{table}
\centering
\begin{tabular}{ |c|c|c| } 
\hline
Field & $SU(3)_c \times SU(2)_L \times U(1)_Y$ & $\mathbb{Z}_3$ \\ 
\hline \hline 
$\phi$ & $(\mathbf{1,2},1/2)$ & $1$\\ \hline  
$L_{e},e_R^{}$ & $(\mathbf{1,2},-1/2)$ & $1$\\ \hline
$L_{\mu},\mu_R^{}$ & $(\mathbf{1,2},-1/2)$ & $\omega$\\ \hline
$L_{\tau},\tau_R^{}$ & $(\mathbf{1,2},-1/2)$ & $\omega^2$ \\
\hline
\end{tabular}
\caption{Quantum numbers of the relevant SM fields under the SM gauge group and $\mathbb{Z}_3$. Here $\omega = \sqrt[3]{-1}$.}
\label{caseA_sm}
\end{table}

\begin{table}
\centering
\begin{tabular}{ |c|c|c| } 
\hline
Field & $SU(3)_c \times SU(2)_L \times U(1)_Y$ & $\mathbb{Z}_3$ \\ 
\hline \hline 
$\Delta$ & $(\mathbf{1,3},2)$ & $1$\\ \hline 
$k^{++}$ & $(\mathbf{1,1},2)$ & $1$\\ \hline 
$k^{+}_e$ & $(\mathbf{1,1},1)$ & $1$\\ \hline 
$k^{+}_\mu$ & $(\mathbf{1,1},1)$ & $\omega$\\ \hline 
$k^{+}_\tau$ & $(\mathbf{1,1},1)$ & $\omega^2$\\ 
\hline
\end{tabular}
\caption{Quantum numbers of the additional fields in Scenario A under the SM gauge group and $\mathbb{Z}_3$.}
\label{caseA_bsm}
\end{table}

The most general renormalizable scalar potential is expressed as the sum of quadratic, trilinear and quartic terms as 
\bea
V &=& V_2 + V_3 + V_4,
\eea
where,
\besub
\bea
V_2 &=& \m^2_{\phi} (\phi^{\dagger} \phi) + M^2_{\Delta} \text{Tr} (\Delta^{\dagger}\Delta) + m^2_k |k^{++}|^2 +  M^2_{\a\b} k^+_\a k^-_\b, 
\\
V_3 &=& \mu_1^{}\, \phi^T (i \sigma_2^{}) \Delta^{\dagger} \phi 
 + \mu_2^{}\, \text{Tr}\big( \Delta^{\dagger} \Delta^{\dagger}\big) k^{++}
 +  \mu_{\a\b}^{}\, k^+_\a k^+_\b k^{--} + \text{H.c.} \\
V_4 &=& \l (\phi^{\dagger} \phi)^2 + \l_1^{} \phi^{\dagger} \phi \text{Tr} (\Delta^{\dagger}\Delta)
+ \l_2^{}  [\text{Tr} (\Delta^{\dagger}\Delta)]^2
 + \l_3^{}  \text{Tr} [(\Delta^{\dagger}\Delta)^2]
+ \l_4^{} \phi^{\dagger} \Delta \Delta^{\dagger} \phi \nonumber \\
&& 
 + \l_5^{} \phi^{\dagger}\phi |k^{++}|^2 + \l_6^{} \text{Tr} (\Delta^{\dagger}\Delta) |k^{++}|^2  + \l_7^{} \big({\tilde{\phi}}^{\dagger} \Delta \phi k^{--} +  \text{H.c.}\big) + \l_8^{} |k^{++}|^4 
\nonumber \\
&&
+ \l_9^{} \phi^{\dagger} \phi k^+_\a k^-_\a + \l_{10}^{} \text{Tr} (\Delta^{\dagger}\Delta) k^+_\a k^-_\a + \l_{11}^{} k^+_\a k^-_\a k^{++} k^{--} \nonumber \\
&&
 + \l_{12}^{} \phi^{\dagger} \Delta^{\dagger} \phi k^+_e
 + \l_{13}^{} k^+_\a k^-_\a k^+_\b k^-_\b . 
\eea 
\eesub
Throughout the text, the indices $\a,\b$ are used to denote the lepton flavors $e,\mu,\tau$ and repeated indices imply summation. 
We point out that some elements of $M^{2}_{\a\b}$ and $\mu_{\a\b}^{}$ break the $\mathbb{Z}_3$ symmetry softly. 
The off-diagonal entries of the dimension-$2$ terms are violent sources of the lepton flavor violation, 
while the dimension-$3$ terms is necessary for realizing observed neutrino mass spectrum, mixings and CP violation. 
Hereafter, we take minimal $\mathbb{Z}_3$ violation hypothesis, where the $\mathbb{Z}_3$ symmetry 
is violated only by the dimension-$3$ terms. A small deviation from this hypothesis will be commented later on.

Following electroweak symmetry breaking (EWSB), $\phi$ and $\Delta$ can be parameterized as
\besub
\bea
\phi &=&
  \begin{pmatrix}
    G^+ \\
    \frac{1}{\sqrt{2}}(v_{\phi}^{} + \phi_0^{} + i G^0)
  \end{pmatrix},
\\
\Delta &=&
  \begin{pmatrix}
    \frac{\delta^{+}}{\sqrt{2}} & \delta^{++} \\
    \frac{1}{\sqrt{2}}(v_{\Delta}^{} + \delta_0^{} + i \delta_1^{}) & -\frac{\delta^{+}}{\sqrt{2}}
  \end{pmatrix},
\eea
\eesub
where $v_{\phi}^{}$ and $v_{\Delta}^{}$ are the VEV's of the scalar doublet and triplet, respectively, with $v^2_{\phi} + 2 v_{\Delta}^2 = (246 ~\text{GeV})^2$.
The presence of the scalar triplet VEV leads to a modified $\rho$ parameter at tree level, {\it i.e.},
$\rho = \left( 1 + \frac{2 v^2_{\Delta}}{v^2_\phi} \right) / \left(1 + \frac{4 v^2_{\Delta}}{v^2_\phi} \right)$. The current bound of $\rho = 1.0004^{+0.0003}_{-0.0004}$~\cite{Patrignani:2016xqp} 
leads to $v_{\Delta}^{} < 5$~GeV.

We now briefly discuss the scalar spectrum of this scenario.  The scalar potential generally allows mixing among the scalar states of the same charge. In terms of mass eigenstates, the neutral scalars in this model are: two $CP$-even scalars ($h,H$) and one $CP$-odd scalar ($A$). The mixing in the neutral sector is therefore identical to the Type-II seesaw model. More details on this part can be found in Refs.~\cite{Dev:2013ff,Aoki:2011pz} and are omitted here for brevity. 
An important impact of the EWSB is the mixing between the two doubly charged states $\delta^{++}$ and $k^{++}$. Diagonalizing the corresponding mass matrix through a rotation by $\theta$ leads to the mass eigenstates $H_1^{++}$ and $H_2^{++}$:
\besub
\bea
\delta^{++} &=& c_{\theta}^{} H_1^{++} + s_{\theta}^{} H_2^{++} 
~, \\
k^{++} &=& -s_{\theta}^{} H_1^{++} + c_{\theta}^{} H_2^{++} 
~.
\eea
\eesub
We also list below the expressions of the $H^{++}_{1,2}$ masses and $\theta$ for $v_\Delta^{} \ll v_\phi^{}$:
\besub
\bea
(M_{1,2}^{++})^2 &=& \frac{1}{2}\big[(A + B) \pm 
\sqrt{(A - B)^2 + 4 C^2}\big]\label{mpp} ~, \\
\text{tan} 2\theta &=& \frac{2 C}{B - A} ~, ~~~\text{where}\label{tan} \\
A &=& M^2_{\Delta} + \frac{1}{2} \l_1^{} v^2 ~,\label{Aprime} \\
B &=& M^2_{k} + \frac{1}{2} \l_5^{} v^2 ~,\label{Bprime} \\
C &=& \frac{1}{2}\l_7^{} v^2 ~.\label{Cprime} 
\eea
\eesub
It follows from Eq.~(\ref{Cprime}) that $\theta \neq 0$ demands $\l_7 \neq 0$.

The next thing taken up is the mixing among the singly charged states. In general, the mixing among 
$\phi^+,\delta^+, k^+_e, k^+_\mu, k^+_\tau$ is governed by a $5 \times 5$ matrix. We, however, shall take the $\l_{12} \rightarrow 0$ limit in this study, as a result of which the $\phi^+$--$\delta^+$
mixing decouples from the remaining 3 $\times$ 3 part. 
This small mixing limit is justified by $v_{\Delta}^{} \ll v_{\phi}^{}$.
In this limit, the $2\times2$ mixing matrix for $\phi^{+}$ and $\delta^{+}$ becomes identical to that in the
pure Type-II seesaw model, giving rise to the Goldstone boson $G^{+}$ and the singly charged physical scalar $H^{+}$ in the mass basis. Since we assume no dimension-2 soft breaking terms in the scalar potential, the 
$3 \times 3$ submatrix spanned by $(k^+_e, k^+_\mu, k^+_\tau)$ is also 
diagonal: diag($M_e^+, M_\m^+, M_\tau^+$). 
Thus, mass eigenstates are the same as in the flavor basis: $H_{\a}^{+}\equiv k_{\a}^{+}$. 

The softly $\mathbb{Z}_3$-violating trilinear interaction then can be recast 
in terms of $H^{++}_i$ ($i = 1,2$) using the mixing angle $\theta$ as
\bea
V_3^{\mathbb{Z}_3 ~\text{breaking}} 
&=& \mu_{\a \b}^{} H^+_{\a} H^+_{\b}
(-\sin\theta H_1^{--} + \cos\theta H_2^{--}) + {\rm H.c.}
\eea

We next discuss the Yukawa Lagrangian in this model.  
The following additional terms are allowed under the $\mathbb{Z}_3$ symmetry:
\bea
\mathcal{L}_{\text{Y}} &=& 
 - y_{\Delta}^{ee}\, \overline{L^{c}_{e}}\, (i\sigma_{2}^{}) \Delta {L_{e}}
  - y_S^{ee}\, \overline{e_{R}^{c}}\, {e}_{R}^{} k_{}^{++}
   - 2\, y_{\Delta}^{\mu \tau}\, \overline{L_{\mu}^{c}}\, i\sigma_{2} \Delta {L_{\tau}}
  - 2\, y_S^{\mu \tau}\, \overline{\mu_{R}^{c}}\, {\tau}_{R}^{} k^{++}  \nonumber \\
&&  
   -  \sum_{\alpha}\, y_A^{\a}\, \epsilon^{\a\b\gamma}\, \overline{L_{\b}^{c}}\, i\sigma_{2} L_{\gamma} k^+_{\a}
   + \rm H.c \label{Yukawalag_A}
\eea
Fermionic statistics demands $y_{\Delta}^{\a \b} = y_{\Delta}^{\b \a}$ and $y_{S}^{\a \b} = y_{S}^{\b \a}$.  A combinatorial factor of $2$ shows up in Eq.~(\ref{Yukawalag_A}). We note that apart from the $(ee)$ and $(\mu \tau)$ elements, $\langle \Delta \rangle$ does not contribute to the other elements of the neutrino mass matrix.
The matrices that describe the Yukawa interactions of $H^+_\a$ and consistent with the $\mathbb{Z}_3$ symmetry are
\bea
y_A^{k_e} =
  \begin{pmatrix}
   0 & 0 & 0 \\
 0 & 0 &  y_A^{e} \\
  0 & -y_A^{e} & 0
  \end{pmatrix},
~y_A^{k_\mu} =
  \begin{pmatrix}
   0 & 0 & -y_A^{\mu} \\
 0 & 0 & 0 \\
  y_A^{\mu} & 0 & 0
  \end{pmatrix},
~y_A^{k_\tau} =
  \begin{pmatrix}
   0 & y_A^{\tau} & 0 \\
 -y_A^{\tau} & 0 & 0 \\
  0 & 0 & 0
  \end{pmatrix}.
\eea
All parameters apart from $\mu_{\a \b}^{}$ are henceforth taken to be real in this scenario.

\subsection{Scenario B: Type-II seesaw realization
}\label{model_B}

The additional scalars introduced in this scenario are three $SU(2)_L$ triplets, $\Delta_e,\Delta_\mu,\Delta_\tau$ and one doubly charged singlet, $k_{}^{++}$. Once again, a softly broken $\mathbb{Z}_3$ symmetry is imposed and 
the charge assignment is given in Table~\ref{caseB}. 
Those of the Higgs doublet and the SM leptons are the same as in Table \ref{caseA_sm}. 
Note that the number of new multiplets in Scenario B is smaller than that in Scenario A, 
which makes the model more restrictive, whereas the number of new particles in Scenario A is smaller.

\begin{table}[htb]
\centering
\begin{tabular}{|c||c|c|}
\hline
Field & $SU(3)_C \times SU(2)_L \times U(1)_Y$ & $\mathbb{Z}_3$\\
\hline\hline
$k_{}^{++}$ & ($\mathbf{1,1},2$) & $1$\\
\hline
$\Delta_e$ & ($\mathbf{1,3},1$) & $1$\\
\hline
$\Delta_\mu$ & ($\mathbf{1,3},1$) & $\omega$\\
\hline
$\Delta_\tau$ & ($\mathbf{1,3},1$) & $\omega^2$\\
\hline
\end{tabular}
\caption{Quantum numbers of the additional scalar fields in Scenario B under the SM gauge group and $\mathbb{Z}_3$. }
\label{caseB}
\end{table}

The scalar potential reads:
\begin{align}
V = V_2 + V_3 + V_4,
\end{align}
with
\begin{align}
V_2
=&
\mu_\phi^2(\phi^\dagger\phi)
+ M_{\Delta\a\b}^2
\mathrm{Tr}(\Delta_\a^\dagger\Delta_\b)
+M_k^2|k^{++}|^2 , \\
V_3
=&
\mu_{e}^{}\, \phi^{\mathrm{T}}(i\sigma_2)
\Delta_e^\dagger\phi
+\mu_{\mu}^{}\, \phi^{\mathrm{T}}(i\sigma_2)
\Delta_\mu^\dagger\phi+
+\mu_{\tau}^{}\, \phi^{\mathrm{T}}(i\sigma_2)
\Delta_\tau^\dagger\phi]\nonumber\\
&
+\mu_{2}^{}\,
k^{++}\mathrm{Tr}(\Delta_e^\dagger \Delta_e^\dagger)
+\mathrm{H.c.} , \\
V_4
=&
\lambda(\phi^\dagger\phi)^2
+\lambda_{1\alpha}
(\phi^\dagger\phi)
\mathrm{Tr}
(\Delta_\alpha^\dagger\Delta_\alpha)\nonumber\\
&+(\lambda_{2\alpha\beta\gamma\delta}
\mathrm{Tr}(\Delta_\alpha^\dagger\Delta_\beta)
\mathrm{Tr}(\Delta_\gamma^\dagger\Delta_\delta)
+\mathrm{H.c.})\nonumber\\
&+(\lambda_{3\alpha\beta\gamma\delta}
\mathrm{Tr}
(\Delta_\alpha^\dagger\Delta_\beta
\Delta^\dagger_\gamma\Delta_\delta)
+\mathrm{H.c.})\nonumber\\
&+\lambda_{4\alpha}
\phi^\dagger\Delta_\alpha\Delta_\alpha^\dagger\phi
+\lambda_5\phi^\dagger\phi|k^{++}|^2
+\lambda_{6\alpha}\mathrm{Tr}
(\Delta_\alpha^\dagger\Delta_\alpha)
|k^{++}|^2\nonumber\\
&+\lambda_{7}
(\tilde{\phi}^\dagger\Delta_e\phi k^{--}+\mathrm{H.c.})
+\lambda_8|k^{++}|^4.
\end{align}

We again adopt the minimal $\mathbb{Z}_3$ violation hypothesis, 
where the dimension-2 terms respect the $\mathbb{Z}_3$ symmetry.  
The trilinear $\mathbb{Z}_3$-breaking terms with $\mu_\mu, \mu_\tau \neq 0$ are included since they ensure all the triplets acquire VEV's.
We define $v_{\Delta}^2 = v_{e}^2 + v_{\mu}^2 + v_{\tau}^2$, where, $v_{e}, v_{\mu}, v_{\tau}$ denote the VEV's of the three triplets. Each triplet
comprises
\bea
\Delta_\alpha =
  \begin{pmatrix}
    \frac{\delta^{+}_\alpha}{\sqrt{2}} & \delta^{++}_\alpha \\
    \frac{1}{\sqrt{2}}(v_{\alpha}^{} + \delta_{0\alpha}
     + i \delta_{1\alpha}) & -\frac{\delta^{+}_\alpha}{\sqrt{2}}
  \end{pmatrix}
  ~.
\eea
With $3$ singly charged states and 4 doubly charged states, this scenario is more involved in terms of field content than the previous one.  The mass eigenstates $H_{\a}^{+}$ and $H_{1,2,\m,\tau}^{++}$ are admixtures
of the gauge-basis states. However, in the $v_\Delta \ll v_\phi$ limit, the mixings simplify to the following
\bea
\begin{pmatrix} 
G^+ \\ H_e^+ \\ H_\m^+ \\ H_\tau^+
\end{pmatrix} = \begin{pmatrix} 
\phi^+ \\ \delta_e^+ \\ \delta_\mu^+ \\ \delta_\tau^+
\end{pmatrix} ~,~~
\begin{pmatrix} 
k_{}^{++} \\ \delta_e^{++} \\ \delta_\mu^{++} \\ \delta_\tau^{++} 
\end{pmatrix}
= \begin{pmatrix}
-\sin\theta & \cos\theta & 0 & 0 \\
 \cos\theta & \sin\theta & 0 & 0 \\
 0 & 0 & 1 & 0 \\
 0 & 0 & 0 & 1 \\
  \end{pmatrix}
  \begin{pmatrix}
  H_1^{++} \\ H_2^{++} \\ H_\m^{++} \\ H_\tau^{++}
  \end{pmatrix}
  ~. \label{mixing_B}
\eea

Similar to Scenario A, a non-zero $\l_7$ induces mixing in the $\delta_e^{++}$--$k_{}^{++}$ sector. The masses and the mixing angle $\theta$ can be obtained from Eqs.~(\ref{mpp})-(\ref{Cprime}) with 
the indices appropriately replaced. The masses of the remaining scalars 
in the $v_{\Delta} \ll v_\phi$ limit are given by
\besub
\bea
(M_\a^{+})^2 &=& M^2_{\Delta \alpha} + \frac{1}{4} (\l_{1\a} + 2 \l_{4\a}) v^2, \label{mip} \\
(M_\m^{++})^2 &=& M^2_{\Delta\mu} + \frac{1}{2}
 \l_{1\mu} v^2, \label{m3pp} \\
(M_\tau^{++})^2 &=& M^2_{\Delta\tau} + \frac{1}{2} \l_{1\tau}v^2. \label{m4pp}
\eea
\eesub

The $\mathbb{Z}_3$-governed Yukawa Lagrangian is expanded in the flavor basis as
\bea
\mathcal{L}_{\text{Y}} &=& 
  - y_{\Delta}^{e e}\, \overline{L_{e}^{c}}\, (i\sigma_{2}^{}) \Delta_e {L_{e}}
  - y_S^{e e}\, \overline{e_{R}^{c}}\, {e}_{R}^{} k_{}^{++}
  - 2\, y_{\Delta}^{\mu \tau}\, \overline{L_{\mu}^{c}}\, i\sigma_{2} \Delta_e {L_{\tau}}
  - 2\, y_S^{\mu \tau}\, \overline{\mu_{R}^{c}}\, {\tau}_{R} k_{}^{++}
  \nonumber \\
&&  
  - y_{\Delta}^{\mu \mu}\, \overline{L_{\mu}^{c}}\, (i\sigma_{2}^{}) \Delta_\mu {L_{\mu}}
  - 2\, y_{\Delta}^{e \tau}\, \overline{L_{e}^{c}}\, (i\sigma_{2}^{}) \Delta_\mu {L_{\tau}} 
  - y_{\Delta}^{\tau \tau}\, \overline{L_{\tau}^{c}}\, (i\sigma_{2}^{}) \Delta_\tau 
  {L_{\tau}}
  - 2\, y_{\Delta}^{e \mu}\, \overline{L_{e}^{c}}\, (i\sigma_{2}^{}) \Delta_\tau {L_{\mu}}
  \nonumber \\
  &&
   + \rm H.c. \label{Yukawalag_B}
\eea
The Yukawa couplings with the triplets entering Eq.~(\ref{Yukawalag_B}) are taken to be complex. These interactions in the gauge basis for the scalars can therefore be described by the following symmetric matrices:
\begin{align}
y_S^{}=&
\begin{pmatrix}
y_S^{ee}&0&0\\
0&0&y_S^{\mu\tau}\\
0&y_S^{\mu\tau}&0
\end{pmatrix} ~,~~
y_{e\Delta}^{}=
\begin{pmatrix}
y_\Delta^{ee}&0&0\\
0&0&y_\Delta^{\mu\tau}\\
0&y_\Delta^{\mu\tau}&0
\end{pmatrix} ~, \nonumber \\
y_{\mu\Delta}^{}=&
\begin{pmatrix}
0&0&y_\Delta^{e\tau}\\
0&y_\Delta^{\mu\mu}&0\\
y_\Delta^{e\tau}&0&0
\end{pmatrix} ~,~~
y_{\tau\Delta}^{}=
\begin{pmatrix}
0&y_\Delta^{e\mu}&0\\
y_\Delta^{e\mu}&0&0\\
0&0&y_\Delta^{\tau\tau}
\end{pmatrix} ~.
\end{align}

Before closing this section, we give the neutrino mass matrix as follows:
\bea
m_\nu = \sqrt{2}
  \begin{pmatrix}
   y_{\Delta}^{ee} v_e & y_{\Delta}^{e\mu} v_\tau & y_{\Delta}^{e \tau} v_\mu \\
 y_{\Delta}^{e\mu} v_\tau & y_{\Delta}^{\mu \mu} v_\mu & y_{\Delta}^{\mu \tau} v_e \\
  y_{\Delta}^{e \tau} v_\mu & y_{\Delta}^{\mu \tau} v_e & y_{\Delta}^{\tau \tau} v_\tau
  \end{pmatrix}
  ~. \label{numass_B}
\eea
We note in passing that the generation of a realistic neutrino mass matrix through the tree-level Type-II fashion demands that each triplet has a VEV (see Eq.~(\ref{numass_B})). This therefore makes it compulsory to include the dimension-3 soft breaking terms.

\section{Numerical results: Scenario A}

The numerical analysis corresponding to Scenario A is presented in this section. It is further split in two subsections for convenience.

\subsection{Muon $g-2$ and lepton flavor violation}\label{muon_g-2}

In this section, we discuss the contribution of this model to the muon anomalous magnetic moment and its possible implications on various lepton flavor-violating (LFV) processes. The total muon anomalous magnetic moment, $\Delta a_{\mu}$, is split into individual contributions coming from the various singly charged as well as doubly charged scalars (see Refs.~\cite{PhysRevD.31.105,Lindner:2016bgg} for the relevant formulae) as
\bea
\Delta a_{\mu} = \Delta a^{\Delta^+}_{\mu}
 + \Delta a^{k^{+}}_{\mu}
 + \sum_{i=1,2}\Delta a^{H_i^{++}}_{\mu}\label{amu_A}
 ~,
\eea
where 

\besub
\bea
\Delta a^{\Delta^+}_{\mu} 
&=& -\frac{ m_{\mu}^2}{8 \pi^2 (1 + 2 v^2_{\Delta}/v^2_{\phi})} 
(y^{\mu \tau}_{\Delta})^2 
\int_0^1 dx \frac{x(1-x)}{M^2_{H^+} -m^2_{\mu}(1-x)} \label{amu_Hp_A}
~, \\
\Delta a^{k^+}_{\mu} 
&=& -\frac{m_{\mu}^2}{16 \pi^2}
\sum_{\a=e, \tau} (y_A^{\a})^2 
\int_0^1 dx \frac{x(1-x)}{ (M_{\a}^{+})^2 -m^2_{\mu}(1-x)}, \label{amu_Hip_A} \\
\nonumber \\
\Delta a^{H_i^{++}}_{\mu} 
&=& -\frac{m_{\mu}^2}{4 \pi^2} 
\int_0^1 dx ~x^2 \frac{[(y^{\mu \tau}_{iL})^2 + (y^{\mu \tau}_{iR})^2] 
(1- x) + 2\, y^{\mu \tau}_{iL} y^{\mu \tau}_{iR} (m_{\tau}/m_{\mu})}
{m^2_{\mu} x^2 + (m^2_{\tau} - m^2_{\mu})x + (M^{++}_i)^2  (1 -x)}  \nonumber \\
&&
- ~\frac{m^2_{\mu}}{2 \pi^2} 
 \int_0^1 dx ~x(1-x)\frac{[(y^{\mu \tau}_{iL})^2 + (y^{\mu \tau}_{iR})^2] x + 2\, y^{\mu \tau}_{iL} y^{\mu \tau}_{iR} 
(m_{\tau}/m_{\mu}) }{m^2_{\mu} x^2
  + ((M^{++}_i)^2 - m^2_{\mu})x + m^2_{\tau} (1 -x)} , 
  \label{amu_Hpp_A} 
\eea
\eesub
with 
\besub 
\bea
y^{\a \b}_{1L} &=& y_{\Delta}^{\a \b} c_{\theta}, \\
y^{\a \b}_{1R} &=& y_S^{\a \b} s_{\theta} ~, \\
y^{\a \b}_{2L} &=& y_{\Delta}^{\a \b} s_{\theta}~, \\
y^{\a \b}_{2R} &=& -y_S^{\a \b} c_{\theta} ~.
\eea
\eesub
In the above expressions, $y_{iL}^{}$ and $y_{iR}^{}$ respectively parameterize the left- and right-chiral Yukawa couplings of $H^{++}_i$ as appearing in the Yukawa Lagrangian below:
\bea
\mathcal{L}_Y \subset \sum_i \overline{\ell_{\a}^c} (y_{iL}^{\a\b} P_L + y_{iR}^{\a\b} P_R) \ell_{\b}\, H_i^{++} + \text{H.c.}
\eea
Analytical forms of the various integrals in Eq.~(\ref{amu_Hpp_A}) are given in the Appendix.

According to Eqs. (\ref{amu_Hp_A}) and (\ref{amu_Hip_A}), the singly charged scalar contribution is always negative. 
On the other hand, an inspection of 
Eq.~(\ref{amu_Hpp_A}) shows that a non-zero mixing between $\delta^{++}$ and $k^{++}$ can render a positive contribution through the chirality flipping effect that is proportional to 
$\mathcal{O}({m_\tau} / {m_\mu})$. Hence, it becomes possible to address the muon $g-2$ anomaly in this model through an appropriate choice of the relevant parameters.


Since couplings of the doubly charged scalars to dilepton states other than
$\mu \tau$ and $e e$ are absent in this case, the only LFV process mediated by the doubly charged scalars at tree level is 
$\tau \rightarrow \bar{\mu} e e$. 
This is in contrast to the pure Type-II and Zee-Babu models, where the other tree-level LFV modes
are also allowed.


\bea
\frac{{\rm BR}_{\tau \r \bar{\mu} e e}}{{\rm BR}_{\tau \r \mu \nu \nu}}
&=& 
\frac{1}{4G_{F}^{2}} \Big\{ 
\big( |y_{S}^{\tau\mu}|^{2}|y_{\Delta}^{ee}|^{2}+|y_{\Delta}^{\tau\mu}|^{2}|y_S^{ee}|^{2} \big) 
s_{\theta}^{2} c_{\theta}^{2} 
\Big( \frac1{(M_{1}^{++})^{2}} - \frac1{(M_{2}^{++})^{2}}\Big)^{2} \nonumber \\
&&
+|y_S^{\tau\mu}|^{2}|y_S^{ee}|^{2}
\Big( \frac{s_{\theta}^{2}}{(M_{1}^{++})^{2}} + \frac{c_{\theta}^{2}}{(M_{2}^{++})^{2}}\Big)^{2} \nonumber \\
&&
+|y_\Delta^{\tau\mu}|^{2}|y_\Delta^{ee}|^{2}
\Big( \frac{c_{\theta}^{2}}{(M_{1}^{++})^{2}} + \frac{s_{\theta}^{2}}{(M_{2}^{++})^{2}}\Big)^{2} \Big\}~,
\label{BR_taumuee_A} 
\eea
where $G_F = 1.17 \times 10^{-5} ~\text{GeV}^{-2}$ refers 
to the Fermi coupling constant.
The experimental upper limits on the various LFV processes are summarized in Table~\ref{lfv_bound}. If non-zero mixing $M_{\a\b}^{2}\, (\a\neq\b)$ among the singly charged states $H_\a^+$ is allowed,  
non-vanishing rates of the radiatively driven LFV processes appear, namely, $\mu \rightarrow e \gamma, \tau \rightarrow e \gamma$ and $\tau \rightarrow \mu \gamma$. 

\begin{table}
\centering
\begin{tabular}{ |c|c| } 
\hline
 LFV channel & Experimental bound \\ 
 \hline \hline 
  $\mu \rightarrow e \gamma$ & $<$ 4.2 $\times 10^{-13}$ 
 ~\cite{TheMEG:2016wtm} \\ \hline
   $\tau \rightarrow e \gamma$ & $<$ 1.5 $\times 10^{-8}$ 
 ~\cite{Aubert:2009ag} \\ \hline
   $\tau \rightarrow \mu \gamma$ & $<$ 1.5 $\times 10^{-8}$ 
 ~\cite{Aubert:2009ag} \\ \hline
  $\mu \rightarrow \bar{e} e e$ & $<$ 1 $\times 10^{-12}$ 
 ~\cite{Bellgardt:1987du} \\ \hline
  $\tau \rightarrow \bar{e} e e$ & $<$ 2.7 $\times 10^{-8}$ 
 ~\cite{Hayasaka:2010np} \\ \hline
 $\tau \rightarrow \bar{\mu} e e$ & $<$ 1.5 $\times 10^{-8}$ 
 ~\cite{Hayasaka:2010np} \\ \hline
  $\tau \rightarrow \bar{\mu} e \mu$ & $<$ 2.7 $\times 10^{-8}$ 
 ~\cite{Hayasaka:2010np} \\ \hline
  $\tau \rightarrow \bar{e} \mu \mu$ & $<$ 1.7 $\times 10^{-8}$ 
 ~\cite{Hayasaka:2010np} \\ \hline
   $\tau \rightarrow \bar{e} \mu e$ & $<$ 1.8 $\times 10^{-8}$ 
 ~\cite{Hayasaka:2010np} \\ \hline
   $\tau \rightarrow \bar{\mu} \mu \mu$ & $<$ 2.1 $\times 10^{-8}$ 
 ~\cite{Hayasaka:2010np} \\ \hline
\end{tabular}
\caption{Latest upper limits on LFV branching ratios.}
\label{lfv_bound}
\end{table}

For the subsequent numerical study, we choose the following set of parameters: \\
($v_{\Delta}^{}, M_{H^+}, M_i^{++}, M_\a^+, y_{A}^{\a}, y_{\Delta}^{e e}, y_S^{e e},
 y_{\Delta}^{\mu \tau}, y_{S}^{\mu \tau}, \theta$) as the basis of independent parameters.
For the numerical analysis, we define $\Delta M = M_2^{++} - M_1^{++}$ and
make the representative choices of $\Delta M = 10, ~50, ~100$~GeV and $\theta = \frac{\pi}{4}, ~\frac{\pi}{10}$ to reduce computational time. The following scan is made:
\besub
\bea
500 ~\text{GeV} < M_1^{++} < 5 ~\text{TeV} ~, \\
-\sqrt{4 \pi} < y_S^{\mu \tau}, y_{\Delta}^{\mu \tau}, y_S^{e e}, 
y_{\Delta}^{e e}, y_A^{e}, y_A^{\mu}, y_A^{\tau} < \sqrt{4 \pi} ~.
\eea
\eesub 
We choose $M_{H^+}^{} = M_1^{++}$ in this analysis for simplicity. The other parameters are fixed as $M^+_{e} = 810$~GeV, $M^+_{\m} = 800$~GeV,  $M^+_{\tau} = 820$~GeV, and $v_{\Delta}^{}= 10^{-15}$~GeV. 
\footnote{This value of the triplet VEV contributes to the neutrino mass elements negligibly. The principal contribution is generated radiatively as will be discussed in the next section.} The singly charged scalar masses are not constrained from LFV in this model. The scalars $k_\a^+$ still contribute to $h \rightarrow \gamma \gamma$. However, 
the $h$--$k_\a^+$--$k_\a^-$ coupling is given by a linear combination of $\l_9^{}$ and $\l_{10}^{}$, and, these quartic couplings do not appear in the rest of the analysis. The
contribution to $h \rightarrow \gamma \gamma$ amplitude from the $k_\a^+$ loops is therefore rendered negligible by
choosing small $\l_9,\l_{10}$ without having to make $k_\a^+$ too heavy.
We still adhere to the aforementioned conservative bound of $\simeq$ 800 GeV keeping in mind possible direct search constraints.

Model points are randomly generated in the aforementioned ranges and tested by the following constraints: 
\begin{enumerate}

\item The muon $g-2$ is within its 2$\,\sigma$ interval, {\it i.e.}, 
$12 \times 10^{-10} \leq \Delta a_\mu \leq 44 \times 10^{-10}$.

\item The LFV processes remain within their respective bounds.

\item The quartic coupling $\l_7 = 2\, s_{\theta}\, c_{\theta} \left[ (M_2^{++})^2 - (M_1^{++})^2 \right]/{v^2}$ remains perturbative, {\it i.e.}, $|\l_7|$ $\leq 4 \pi$

\end{enumerate}
Points clearing the constraints are then kept and used in the following analysis.

\begin{figure}[tbhp]
\begin{center}
%
\includegraphics[scale=0.50]{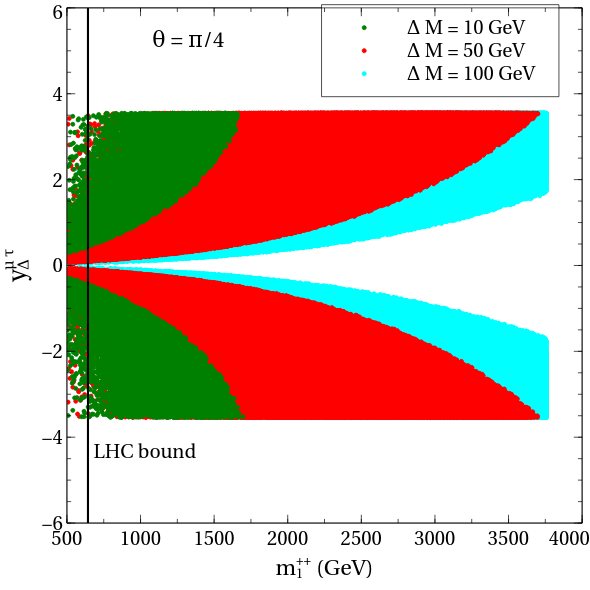}~~~ 
\includegraphics[scale=0.50]{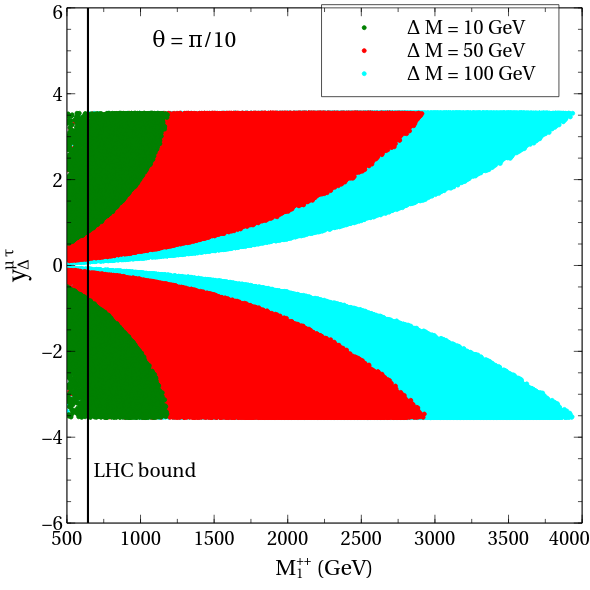}~~~\\ 
\caption{The allowed parameter space maintaining $\Delta a_\mu$ 
within its 2$\sigma$ range in the $M_1^{++} - y_{\Delta}^{\mu \tau}$
plane for $\theta = \frac{\pi}{4}$ (left) and $\frac{\pi}{10}$ (right). The color coding is explained in the legends. The $M_1^{++}$ values left to the vertical line are disallowed by like-sign dilepton searches at the LHC~\cite{CMS:2017pet}.}
\label{f:ydelmutau-mH1pp}
\end{center}
\end{figure} 

Fig.~\ref{f:ydelmutau-mH1pp} shows the allowed parameter space in the $M_1^{++}$--$y_\Delta^{\mu\tau}$ plane for the choices of $\Delta M = 10, ~50, ~100$~GeV in green, red and cyan, respectively and for $\theta = \frac{\pi}{4}$ (left plot) and $\frac{\pi}{10}$ (right plot).
The contribution to $\Delta a_{\mu}$ from $H^+$ is roughly given by 
$\frac{-(y_\Delta^{\mu \tau})^2}{48 \pi^2} \frac{m^2_\mu}{M_{H^{+}}^2}$, 
whereas the chirality flipping term from $H^{++}$ in this model has 
\bea
\Delta a_\mu  \simeq
 \frac{y_{\Delta}^{\mu \tau} y_{S}^{\mu \tau}}{16 \pi^2} \frac{m_\mu m_{\tau}}{(M_1^{++})^3} \Delta M
 s_{\theta} c_{\theta} \log\frac{m^2_\tau}{(M_1^{++})^2}
  ~.
\eea 

In comparison, the singly charged contribution is suppressed by roughly a factor of $ m_\tau \Delta M/m^2_\mu$ for $M_{H^{+}}^{} \simeq M_1^{++}$
and $y_S^{\mu \tau} \simeq y_\Delta^{\mu \tau}$, thereby rendering the doubly charged scalars the dominant contributors. Two crucial parameters in this case are therefore $\theta$ and $\Delta M$. A higher $\Delta M$ implies a larger positive contribution to $\Delta a_\mu$. For a fixed value of $\Delta a_\mu$, a higher value of $\Delta M$ also implies a higher maximally allowed value for $M_1^{++}$. For example, the left plot in Fig.~\ref{f:ydelmutau-mH1pp} shows $M_1^{++} \lesssim 3.7$ TeV in case of $\Delta M = 50$~GeV and $M_1^{++} \lesssim 1.7$ TeV in case of $\Delta M = 10$~GeV. The scalar coupling $\l_7$ hits its perturbative limit for $\Delta M = 100$~GeV and $\theta = \frac{\pi}{4}$, thereby disfavoring $M_1^{++} \gtrsim 3.75$~TeV for this particular choice. This explains the sharp vertical boundary in the left plot. 
In addition, $\theta=\frac{\pi}{4}$ maximizes $\Delta a_{\mu}$ when the other parameters are held fixed. This leads to the expectation that the allowed range of $M_1^{++}$ will be the most relaxed. This is again confirmed in the plots, where for $\Delta M = 50$~GeV, $M_1^{++} \lesssim 3.6$~TeV for $\theta = \frac{\pi}{4}$ while $M_1^{++} \lesssim 2.9$~TeV for $\theta = \frac{\pi}{10}$.

\begin{figure}[tbhp]
\begin{center}
%
\includegraphics[scale=0.50]{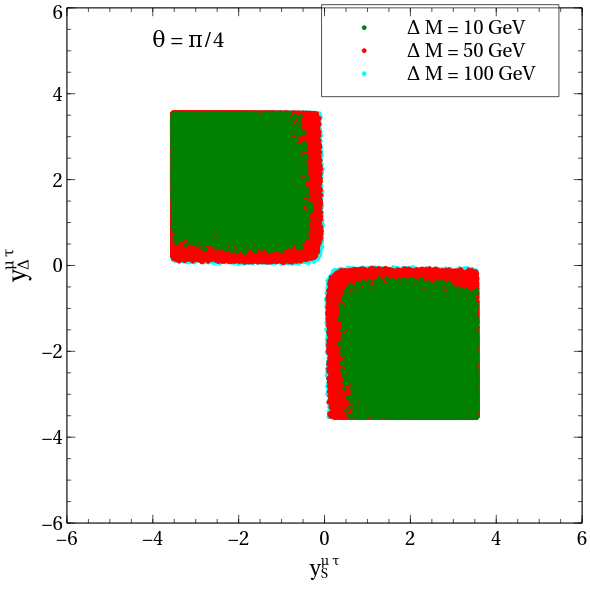}~~~ 
\includegraphics[scale=0.50]{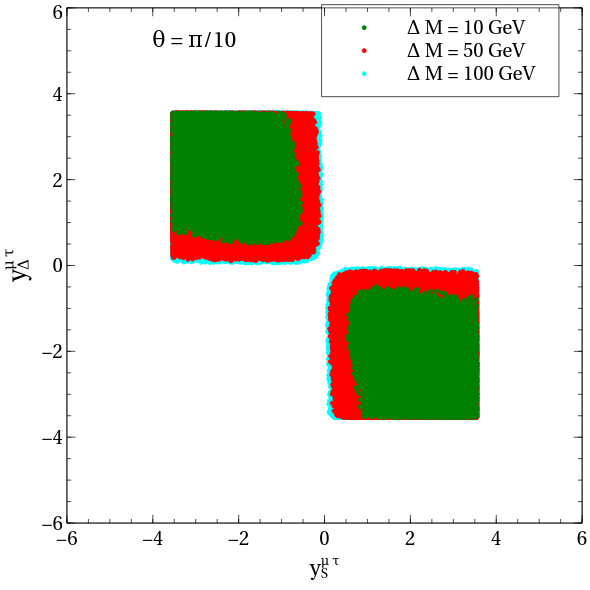}~~~\\ 
\caption{The allowed parameter space maintaining $\Delta a_\mu$ 
within its 2$\sigma$ range in the $y_{\Delta}^{\mu \tau} - y_S^{\mu \tau}$
plane for $\theta = \frac{\pi}{4}$ (left) and $\frac{\pi}{10}$ (right). The color coding is explained in the legends.}
\label{f:ydelmutau-ySmutau}
\end{center}
\end{figure}

The same parameter points are plotted in the $y_{\Delta}^{\mu \tau}$--$y_{S}^{\mu \tau}$ plane in Fig.~\ref{f:ydelmutau-ySmutau}. It is seen that points are distributed along the entire ranges of both Yukawa couplings whenever $\Delta M = 100$~GeV. The same allowed ranges for both couplings can be traced back to the invariance of the chirality flip under $y_\Delta^{\mu \tau} \leftrightarrow y_S^{\mu \tau}$. For lower $\Delta M$ values, low values of the Yukawa couplings tend to be disfavored, albeit the reduction in the parameter space is not appreciable.  Hence, no  strong constraint is imposed by $\Delta a_\mu$ in this parameter space.

\begin{figure}[tbhp]
\begin{center}
%
\includegraphics[scale=0.50]{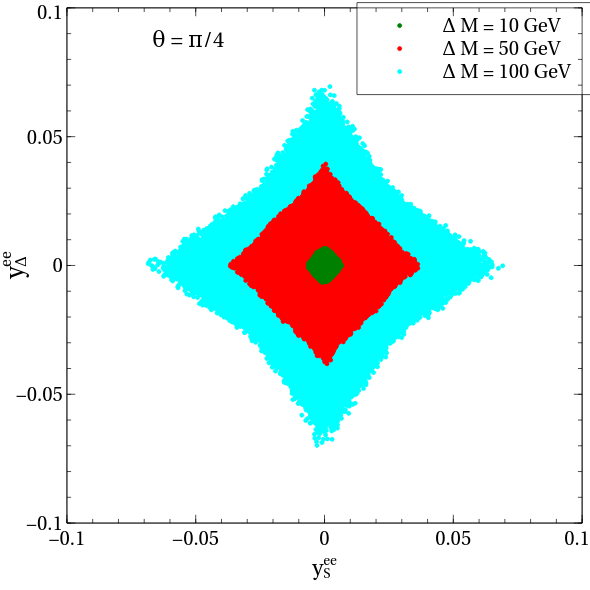}~~~ 
\includegraphics[scale=0.50]{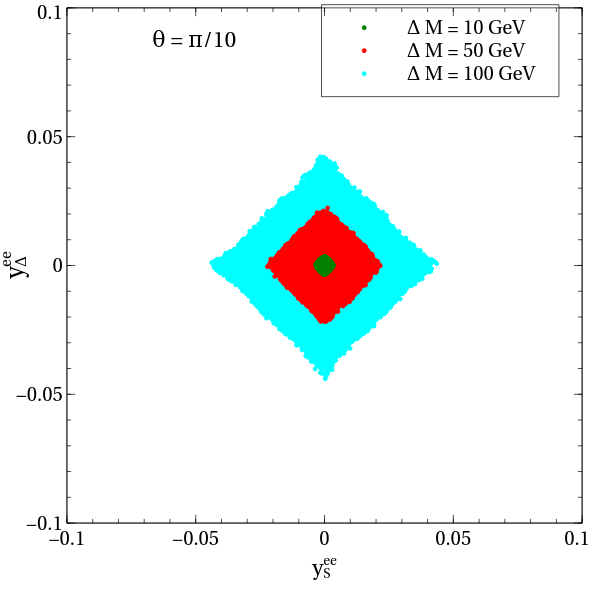}~~~\\ 
\caption{The allowed parameter space maintaining $\Delta a_\mu$ 
within its 2$\sigma$ range in the $y_{\Delta}^{e e} - y_S^{e e}$
plane for $\theta = \frac{\pi}{4}$ (left) and $\frac{\pi}{10}$ (right). The color coding is explained in the legends.}
\label{f:ydelee-ySee}
\end{center}
\end{figure} 

\begin{figure}[tbhp]
\begin{center}
%
\includegraphics[scale=0.50]{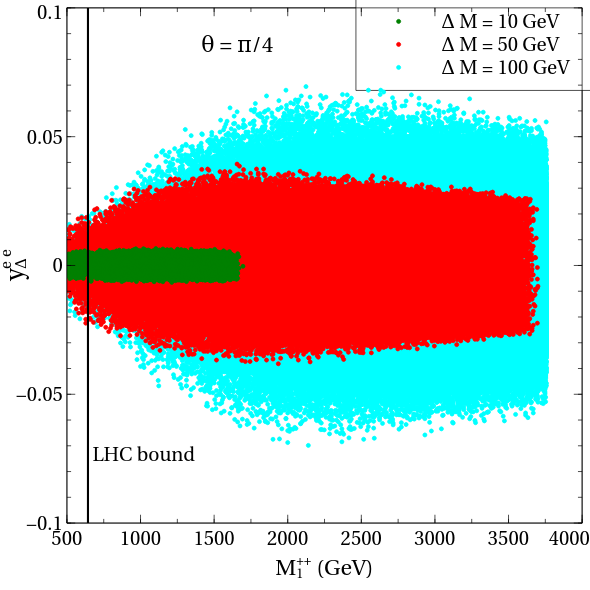}~~~ 
\includegraphics[scale=0.50]{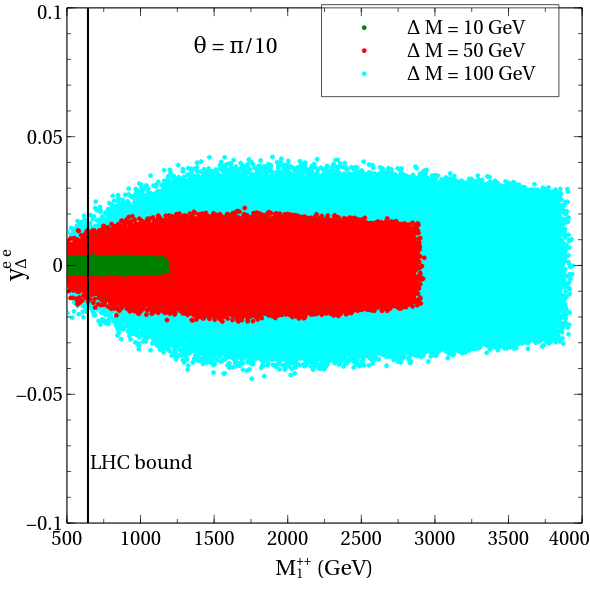}~~~\\ 
\caption{The allowed parameter space maintaining $\Delta a_\mu$ 
within its 2$\sigma$ range in the $y_{\Delta}^{e e} - y_S^{e e}$
plane for $\theta = \frac{\pi}{4}$ (left) and $\frac{\pi}{10}$ (right). The color coding is explained in the legends. The $M_1^{++}$ values left of the vertical line are disallowed by like-sign dilepton searches at the LHC.}
\label{f:m1pp-ydelee}
\end{center}
\end{figure} 

\begin{figure}[tbhp]
\begin{center}
%
\includegraphics[scale=0.50]{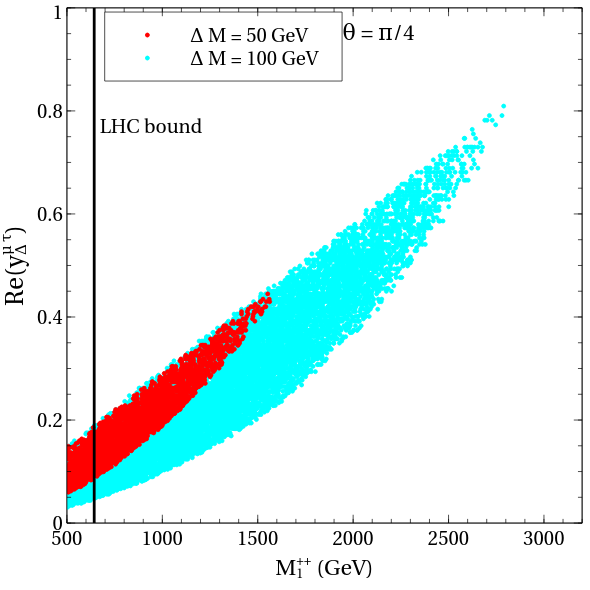}~~~ 
\includegraphics[scale=0.50]{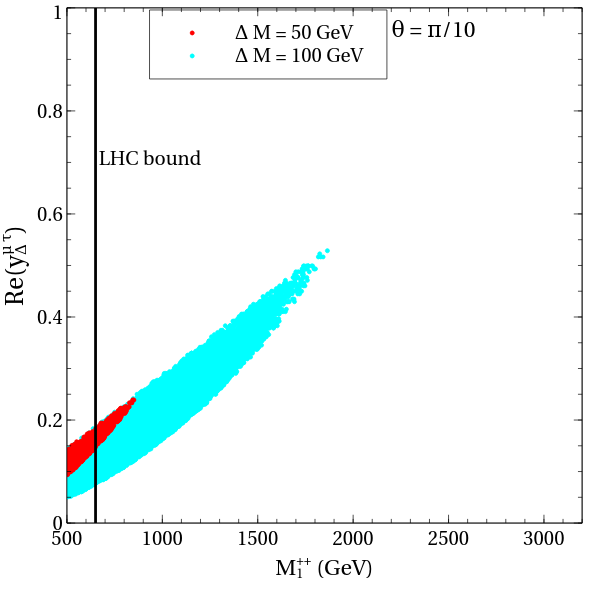}~~~\\ 
\caption{The allowed parameter space maintaining $\Delta a_\mu$ 
within its 2$\sigma$ range and ${\rm BR}_{\tau \rightarrow \bar{\mu} e e}$ within the quoted limit in the $M_1^{++} - {\rm Re}(y_\Delta^{\mu \tau})$ plane for $\theta = \frac{\pi}{4}$ (left) and $\frac{\pi}{10}$ (right). A normal neutrino mass hierarchy is assumed. The color coding is explained in the legends. The region left to the black line is disallowed by the dilepton searches at the LHC.}
\label{f:m1pp-ydelmutau_B}
\end{center}
\end{figure} 

We have taken ${\rm BR}_{\tau \rightarrow \mu \nu \nu} \simeq 1/6$ while
determining ${\rm BR}_{\tau \rightarrow \bar{\mu} e e}$ using Eq.~(\ref{BR_taumuee_A}).
The prediction of the $\tau \rightarrow \bar{\mu} e e$ rate is correlated with that of $\Delta a_\mu$, much due to their dependence on a common set of model parameters, as is evident from Eq.~(\ref{amu_Hpp_A}) and Eq.~(\ref{BR_taumuee_A}). 
Firstly, the allowed range of $|y_{S}^{e e}|$ is similar to that of $|y_{\Delta}^{e e}|$ (see Fig.~\ref{f:ydelee-ySee}). As illustrated in this section, the mass splitting $\Delta M$ becomes crucial in determining the maximum of $M_1^{++}$, from the consideration of $\Delta a_\mu$. And the allowed ranges of $y_{\Delta}^{\mu \tau}$ and $y_{S}^{\mu \tau}$ obviously depend on the overall mass scale of doubly charged scalars. That is, a larger allowed band for $M_1^{++}$ loosens the allowed ranges for $y_{\Delta}^{\mu \tau}$ and $y_{S}^{\mu \tau}$. This is corroborated by an inspection of 
Figs.~\ref{f:ydelee-ySee} and \ref{f:m1pp-ydelee}. 
In the case of $\theta = \frac{\pi}{4}$, $|y_{\Delta}^{e e}| < 10^{-2}$ is obtained for $\Delta M = 10$~GeV while the corresponding bound settles at $\simeq 0.07$ for $\Delta M = 100$~GeV. The bound for $\Delta M = 50$~GeV, as expected, is somewhere in between. The qualitative behavior of the parameter space for other values of $\theta$ and $\Delta M$ can be readily understood from this discussion.

We add here that
the results of the numerical scans presented in this section are not affected by the details in the neutrino sector. This is so because 
a neutrino mass matrix complying with the latest data can always be reconstructed in this model by tuning the trilinear parameters accordingly, as we shall see in the next subsection. The same parameters do not enter the calculations of $\Delta a_\mu^{}$ and the LFV rates.

\subsection{Neutrino mass matrix}\label{numass}

\begin{figure}[h!]
\begin{center}
    \subfloat[ \label{zb1}]{%
      \includegraphics[scale=0.95]{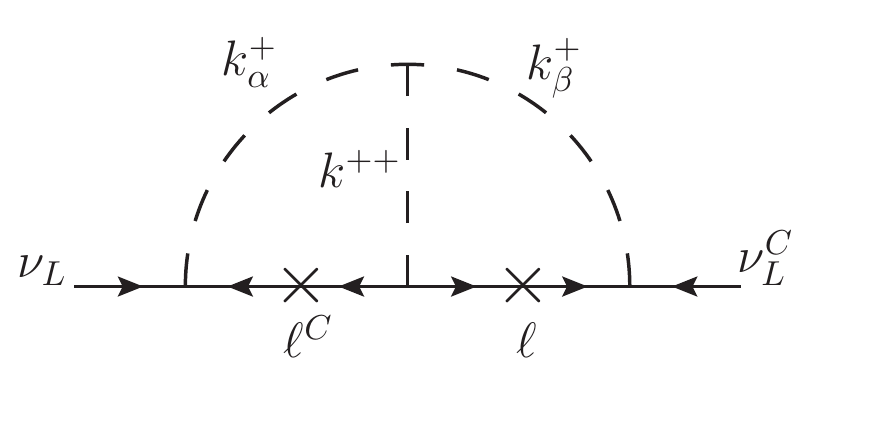}
    }~~~~
    \subfloat[\label{zb2}]{%
      \includegraphics[scale=0.95]{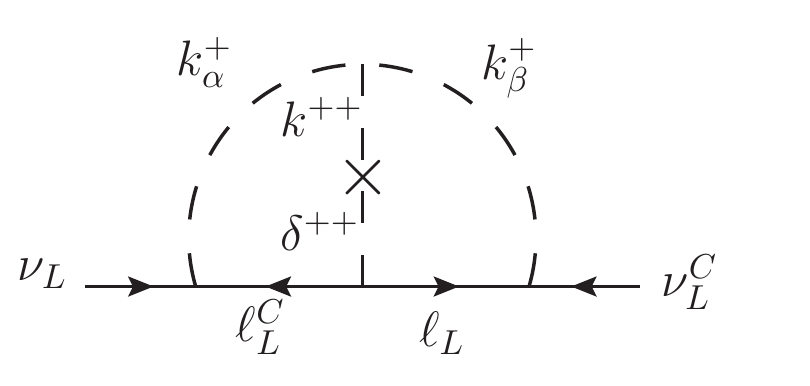}~~~~
    }
\end{center}
\caption{Two-loop graphs responsible for neutrino mass generation.}
\label{f:2loop}
\end{figure}

We discuss details of neutrino mass generation in this section. Similar to what happens in the Zee-Babu model, non-zero mass for the neutrinos arises at the two-loop level in this framework. Representative Feynman graphs are shown in Fig.~\ref{f:2loop}.

We point out here that the amplitude in Fig.~\ref{zb1} is similar to the usual Zee-Babu amplitude as far as its chirality structure is concerned. In contrast, the amplitude in Fig.~\ref{zb2} is induced by the $\delta^{++}$--$k^{++}$ mixing in one of the scalar lines. A different chirality structure
renders it much more enhanced compared to Fig.~\ref{zb1}. 
Explicitly, the neutrino mass matrix elements in this model are given by: 
\bea
m_\nu^{\a\b} &=& \sqrt{2} y_{\Delta}^{\a\b} v_{\Delta} 
\nonumber \\
&& 
- 16\, \sum_{\a'\b'\a''\b''}\, \mu_{\a''\b''}^{}
y_A^{\a''} \epsilon^{\a\a'\a''}  y_A^{\b''} \epsilon^{\b\b'\b''} 
\Big\{ 
y^{\a'\b'}_S \, \Big[s^2_{\theta} I_{k1}^{\a''\b''\a'\b'} + c^2_{\theta} I_{k2}^{\a''\b''\a'\b'} \Big]
\nonumber \\
&& \qquad \qquad \qquad 
+ y^{\a'\b'}_{\Delta} \, s_{\theta} c_{\theta} \, 
\Big[ -I_{\Delta1}^{\a''\b''\a'\b'} + I_{\Delta2}^{\a''\b''\a'\b'} \Big] \Big\}, \label{mv1} 
\eea
where $I_{Xi}^{\a''\b''\a'\b'} \equiv I_X(M^+_{\a''}, M^+_{\b''}, M^{++}_i,m_{\a'},m_{\b'})$. 
The $2$-loop integrals $I_k(M^+_\a, M^+_\b, M^{++}_i, m_{\mu}, m_{\tau})$ and 
$I_\Delta(M^+_\a, M^+_\b, M^{++}_i, m_{\mu}, m_{\tau})$ have been defined and evaluated in the Appendix.  

The $U_{\text{PMNS}}$ matrix diagonalizes the neutrino mass matrix $m_\nu$, {\it i.e.},
\besub
\bea
&&
m_\nu = U_{\text{PMNS}}^* ~m_\nu^{\text{diag}} ~U_{\text{PMNS}}^T
~, \label{nu}\\
&& \text{with}
~U_{\text{PMNS}} = V_{\text{PMNS}} \times 
\text{diag}(1,e^{i \a_{21}/2},,e^{i \a_{31}/2}) ~\mbox{and} \\
&& V_{\text{PMNS}} =  \begin{pmatrix}
    c_{12} c_{13} & s_{12} c_{13} &  s_{13} e^{-i \delta_{CP}}\\
 -s_{12}c_{23} - c_{12}s_{23}s_{13} e^{i \delta_{CP}} & c_{12}c_{23} - s_{12}s_{23}s_{13} e^{i \delta_{CP}} & s_{23}c_{13} \\
  s_{12}s_{23}-c_{12}c_{23}s_{13} e^{i \delta_{CP}} 
  & -c_{12}s_{23} - s_{12}c_{23}s_{13} e^{i \delta_{CP}} & c_{23} c_{13}
  \end{pmatrix}
  ~,
\eea
\eesub
where $s_{ij} =\sin\theta_{ij}$, $c_{ij} =\cos\theta_{ij}$, $\delta_{CP}^{}$ is the Dirac phase, and $\a_{21}^{}$ and $\a_{31}^{}$ are the Majorana phases.

Before proceeding further, a comment on the relative magnitudes of 
$I_k$ and $I_\Delta$ is in order. 
Due to different chirality structures, $I_k/I_\Delta \sim \mathcal{O}(m^2_\ell/M^2_S)$, where $m_\ell^{}$ and
$M_S^{}$ denote a lepton mass and a scalar mass, respectively.  Any contribution from $I_k$ can hence be neglected for this model.
One may refer to Ref.~\cite{McDonald:2003zj} and some references therein to gain additional insight in the two-loop functions.

We have fitted the neutrino oscillation data using our model in the following approach. An $m_\nu^{\a \b}$ has six complex entries that are derivable from the neutrino oscillation parameters (see Eq.~\eqref{nu}).
There are 6 complex $\mu_{\a \b}^{}$ in this model. Each $\mu_{\a \b}^{}$ can therefore be solved for from Eq.~\eqref{mv1}. 
We recall that all the Yukawa couplings are taken to be real and therefore $\mu_{\a \b}^{}$ are necessarily complex in order to account for the phases coming from $\delta_{CP}^{}, \a_{21}^{}$ and $\a_{31}^{}$.

One can make the following order-of-magnitude estimate for $\mu_{\a \b}^{}$. First, let's assume $v_\Delta^{} = 10^{-15}$~GeV so that there is no noticeable contribution from $\langle \Delta \rangle$ to any of the neutrino mass elements.  
Then for $M_\a^+ \simeq 800$~GeV, $M_i^{++} \simeq 1$~TeV, the $I_\Delta^{}$ integral is of $\mathcal{O}(10^{-4})$. Considering a typical $m_\nu^{\a \b}$ having an absolute value around $\mathcal{O}(10^{-3})$~eV and assuming the Yukawa couplings of $\mathcal{O}(1)$, the $\mu_{\a \b}^{}$ value is about
$\mathcal{O}(10^{-8})$~GeV. As expected, this is several orders of magnitude smaller than what it would have been in case where only the Zee-Babu-like amplitude (Fig.~\ref{zb1}) is present. 
Noting that the new $2$-loop amplitude as shown in Fig.~\ref{zb2} survives only in the $\theta \neq 0$ limit, we deem this observation a fallout of the $\Delta$--$k^{++}$ mixing. Therefore, this mixing plays a pivotal role in neutrino mass generation, much like it plays in explaining the muon $g-2$ anomaly.

The full allowed ranges of $\mu_{\a \b}^{}$ can be revealed through a parameter scan. The singly charged scalars are assigned with masses $\simeq$ 800~GeV. Besides, $M_1^{++}$ and the Yukawa couplings are varied in the same ranges as in the previous section.
In addition, the neutrino oscillation parameters are fixed to their central values~\cite{Patrignani:2016xqp} as
\bea
&&
\text{sin}^2\theta_{12} = 0.307 ~,~ 
~\text{sin}^2\theta_{23} = 0.510 ~,~
~\text{sin}^2\theta_{13} = 0.021 ~, \nonumber \\
&&
\Delta m^2_{21}
 = 7.45 \times 10^{-5} ~\text{GeV}^2 ~,
~\Delta m^2_{32} = 2.53 \times 10^{-3} ~\text{GeV}^2 ~,  \nonumber \\
&&
\delta_{CP} =  1.41\pi ~,~ 
~\a_{21} = \a_{31} = 0 ~. \label{nuparam_fixed}
\eea

\begin{figure} 
\begin{center}
%
\includegraphics[scale=0.44]{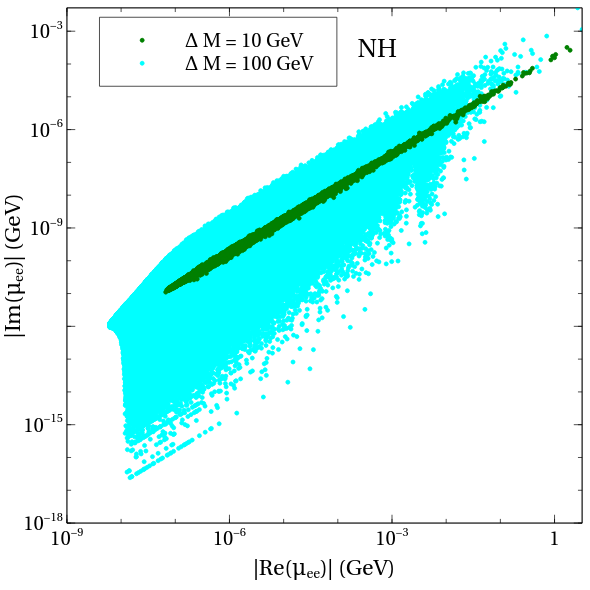}~~~ 
\includegraphics[scale=0.44]{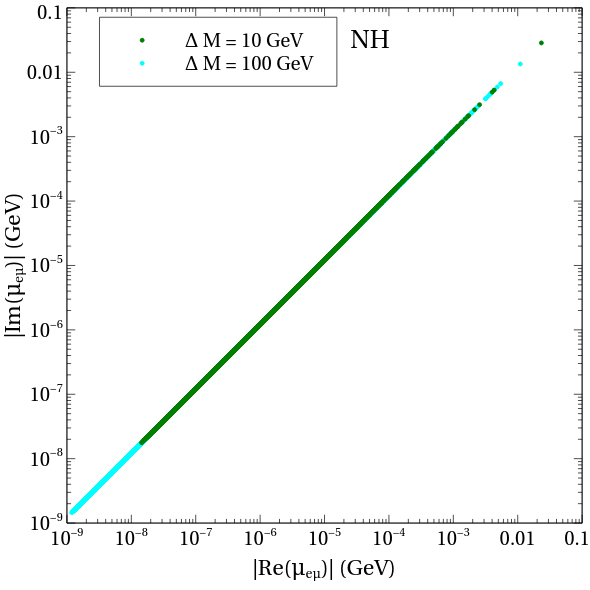}~~~\\ 
\includegraphics[scale=0.44]{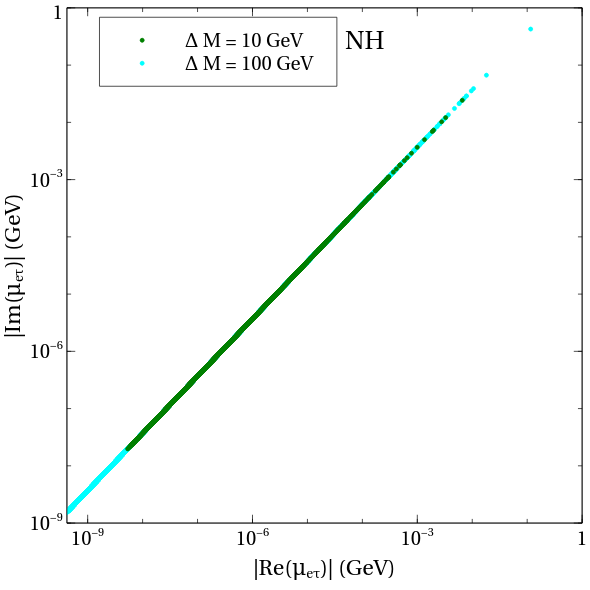}~~~ 
\includegraphics[scale=0.44]{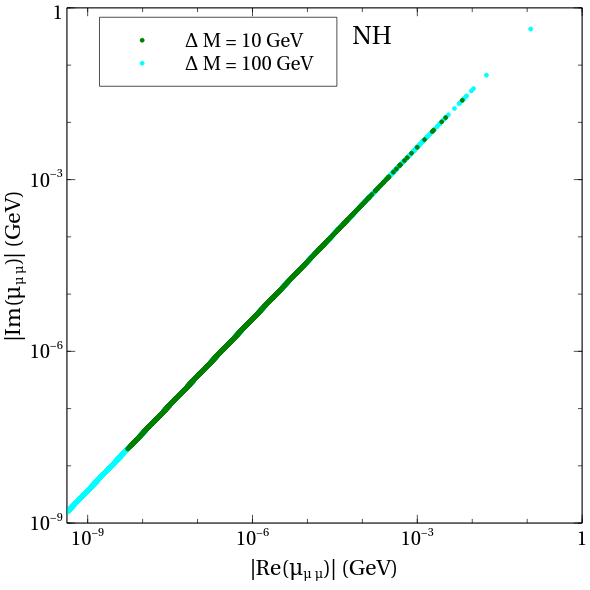}~~~\\
\includegraphics[scale=0.44]{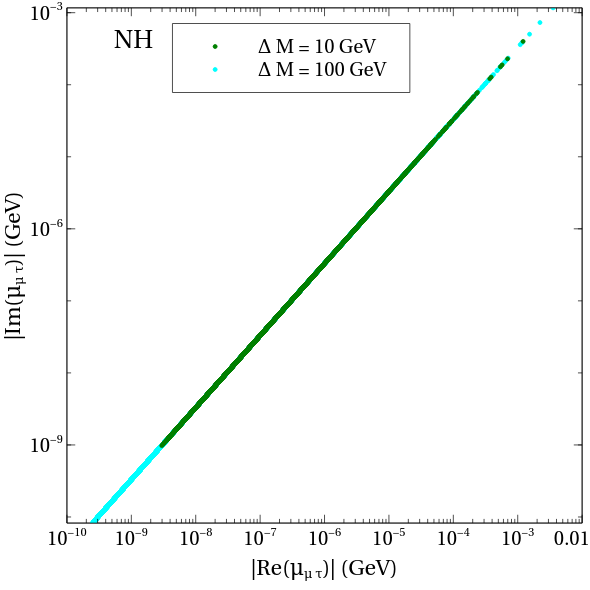}~~~ 
\includegraphics[scale=0.44]{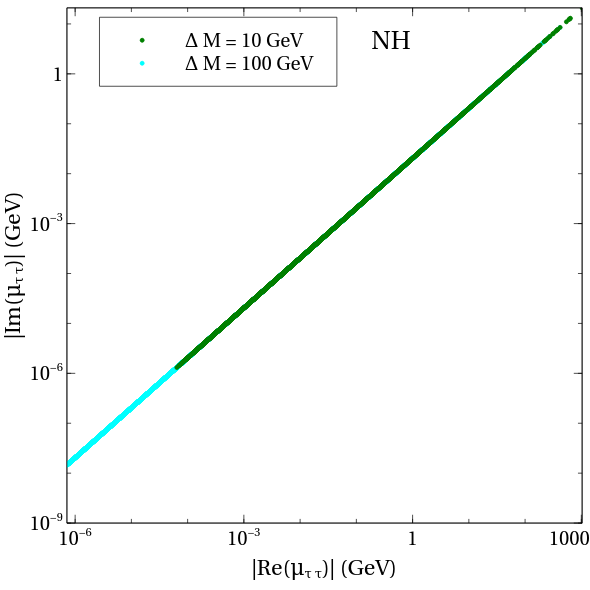}~~~\\
\caption{Allowed values of $\mu_{\a \b}$ in the case of normal hierarchy (NH), plotted in the plane of real vs imaginary axes. The color coding can be read from the legends.}
\label{f:NH}
\end{center}
\end{figure}

\begin{figure} 
\begin{center}
%
\includegraphics[scale=0.44]{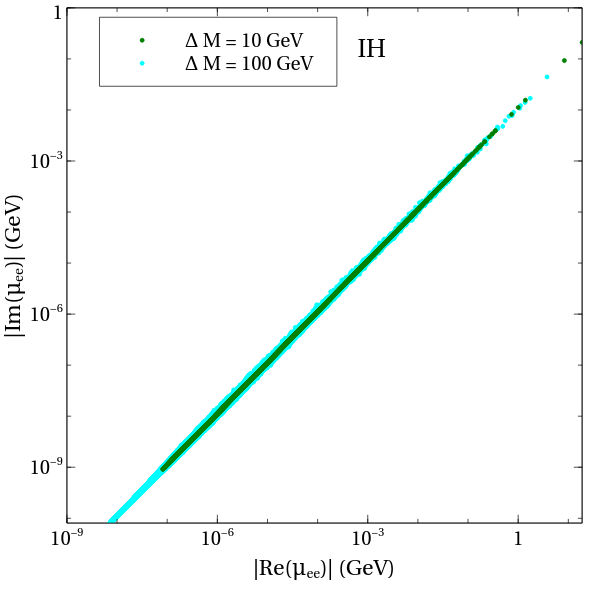}~~~ 
\includegraphics[scale=0.44]{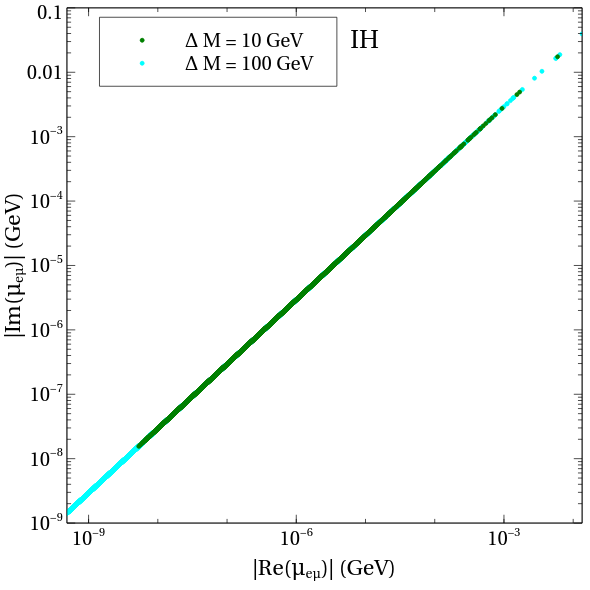}~~~\\ 
\includegraphics[scale=0.44]{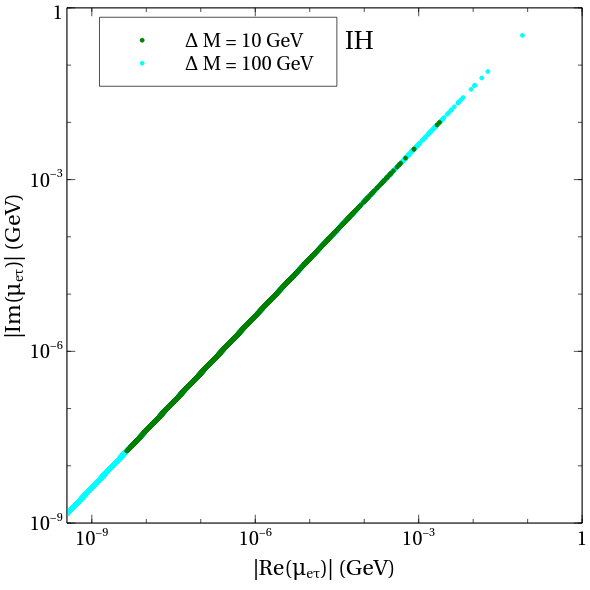}~~~ 
\includegraphics[scale=0.44]{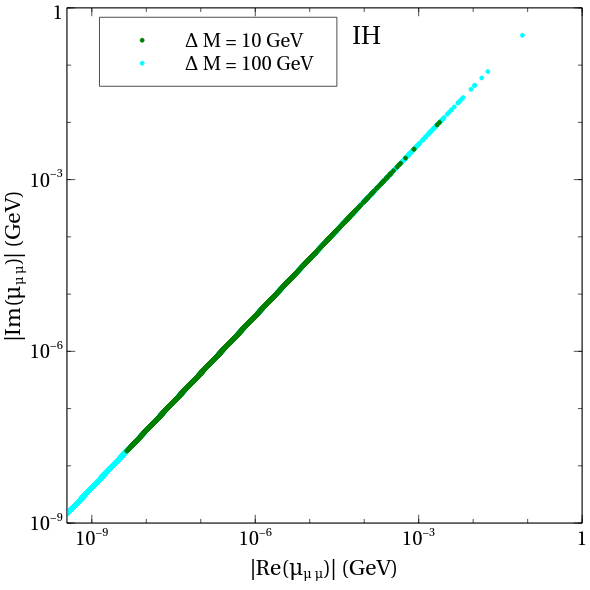}~~~\\
\includegraphics[scale=0.44]{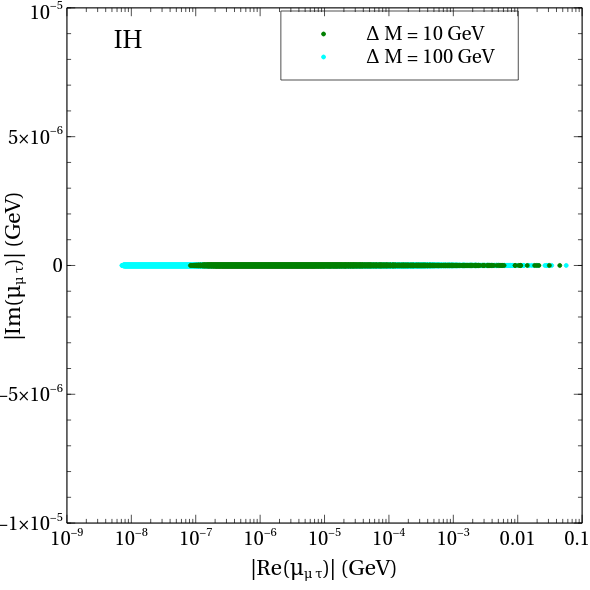}~~~ 
\includegraphics[scale=0.44]{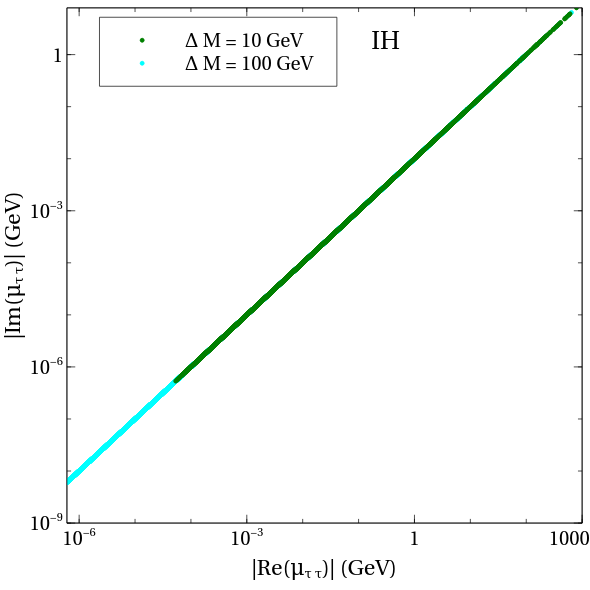}~~~\\
\caption{Allowed values of $\mu_{\a \b}$ in the case of inverted hierarchy (IH), plotted in the plane of real vs imaginary axes. The color coding can be read from the legends.}
\label{f:IH}
\end{center}
\end{figure}

The mass of the lightest neutrino and Majorana phases are assumed to vanish in the present analysis.  
In addition to imposing the constraints of $\Delta a_\mu$, LFV bounds and $|\l_7| < 4 \pi$, we also perform the a perturbativity check of the trilinear parameters, {\it i.e.}, $|\mu_{\a \b}^{}| < 4\pi ~\text{min}(M^{++}_i, M^+_\a)$.  Figs.~\ref{f:NH} and \ref{f:IH} depict the real and imaginary parts of $\mu_{\a\b}^{}$ that are required to explain the neutrino data for normal as well as inverted mass orderings, respectively. In order to understand the linear shape of these plots, consider that $\mu_{\mu\tau}^{} \propto m_\nu^{ee}$, with the proportionality factor being real (since the Yukawa couplings are real). 
One then can write
\bea
\frac{\text{Re}(\mu_{\mu \tau}^{})}{\text{Im}(\mu_{\mu \tau}^{})} &=& 
\frac{\text{Re}(m_\nu^{ee})}{\text{Im}(m_\nu^{ee})}\label{ratio}
\eea
Now, the right hand side of Eq. (\ref{ratio}) is fixed and this in turn fixes the slope of the parameter points in the 
$|\text{Re}(\mu_{\mu \tau}^{})|$--$|\text{Im}(\mu_{\mu \tau}^{})|$ plane. This pattern is also seen in case of trilinear parameters other than 
$\mu_{ee}^{}$. And this difference comes from the fact that the expression for $m_\nu^{\mu \tau}$ constrains contributions from both $\mu_{ee}^{}$ and 
$\mu_{\mu \tau}^{}$.
The linear shape obviously will get smeared once a variation of the neutrino oscillation parameters is invoked.

We comment here that the above analyses can be repeated for a larger value of $v_\Delta^{}$. In such a case, the contribution of the triplet to the $(ee)$ and $(\mu\tau)$ elements can be appreciable and, in fact, much larger than the mass scale of the light neutrinos in principle. If so, $|\mu_{\mu \tau}^{}|$ and $|\mu_{ee}^{}|$ also have to be suitably large so as to make way for a cancellation between the tree-level and two-loop terms. Therefore, no strong constraint on the triplet VEV emerges in this scenario from the consideration of neutrino mass.

\section{Numerical results: Scenario B}\label{results_B}

In this section, we demonstrate the viability of Scenario B in connection to the muon $g-2$ anomaly, neutrino mass and LFV processes. As we will see below, the predictions of $\Delta a_\mu$ and LFV are expected to be sharply correlated with the neutrino masses and mixings for the present scenario. Therefore, we do not divide our discussions into different subsections, as was the approach taken in the case of Scenario A, owing to a different neutrino mass mechanism in that case.
The contributions to muon $g-2$ coming from the singly and doubly charged scalars add up as follows: 
\bea
\Delta a_{\mu} = \Delta a^{\Delta^{+}}_{\mu}
 + \Delta a^{\Delta_{}^{++}}_{\mu}
 + \sum_{i=1,2}\Delta a^{H_i^{++}}_{\mu},\label{amu}
\eea
where
\besub
\bea
\Delta a^{\Delta^+}_{\mu} &=& 
-\frac{m_{\mu}^2}{8 \pi^2} 
 {|y_\Delta^{\mu \tau}|}^2 \int_0^1 dx \frac{x(1-x)}{ (M_e^{+})^2 -m^2_{\mu}(1-x)} \nonumber \\
&&
  -\frac{m_{\mu}^2}{8 \pi^2}
 {|y_\Delta^{\mu \mu}|}^2 \int_0^1 dx \frac{x(1-x)}{ (M_\m^{+})^2 -m^2_{\mu}(1-x)} \nonumber \\
&&
 -\frac{m_{\mu}^2}{8 \pi^2}
 {|y_\Delta^{e \mu}|}^2 \int_0^1 dx \frac{x(1-x)}{ (M_\tau^{+})^2 -m^2_{\mu}(1-x)}, 
\label{amu_Hip} \\ \nonumber \\
\Delta a^{H_i^{++}}_{\mu} &=& -\frac{m_{\mu}^2}{4 \pi^2} 
\int_0^1 dx \frac{(|y^{\mu \tau}_{iL}|^2 + |y^{\mu \tau}_{iR}|^2) 
(1 - x) + 2 ~\text{Re}[y^{\mu \tau}_{iL} y^{\mu \tau}_{iR}] 
(m_{\tau}/m_{\mu})}
{m^2_{\mu} x^2 + (m^2_{\tau} - m^2_{\mu})x + (M^{++}_i)^2  (1 -x)} x^{2} \nonumber \\
&&
- ~\frac{m^2_{\mu}}{2 \pi^2}
 \int_0^1 dx ~\frac{(|y^{\mu \tau}_{iL}|^2 + |y^{\mu \tau}_{iR}|^2) x + 2 ~\text{Re} [y^{\mu \tau}_{iL} y^{\mu \tau}_{iR}] 
(m_{\tau}/m_{\mu})}{m^2_{\mu} x^2
  + ((M^{++}_i)^2 - m^2_{\mu})x + m^2_{\tau} (1 -x)} x(1-x), \label{amu_H12pp} \\
\Delta a^{\Delta_{}^{++}}_{\mu} &=&  
-\frac{m_{\mu}^2 |y_\Delta^{\mu\mu}|^2}{4 \pi^2} 
\int_0^1 dx \frac{x^2(1-x)}{m^2_{\mu} x^2 + (m^2_{\tau} - m^2_{\mu})x + (M^{++}_\m)^2  (1 -x)} \nonumber \\
&&
- ~\frac{m^2_{\mu} |y_\Delta^{\mu\mu}|^2}{2 \pi^2}
 \int_0^1 dx ~\frac{x^2(1-x)}{m^2_{\mu} x^2
  + ((M^{++}_\m)^2 - m^2_{\mu})x + m^2_{\tau} (1 -x)}\label{amu_H3pp} 
 \\
\nonumber \\  
&&  -\frac{m_{\mu}^2 |y_\Delta^{e\mu}|^2}{4 \pi^2} 
\int_0^1 dx \frac{x^2(1-x)}{m^2_{\mu} x^2 + (m^2_{\tau} - m^2_{\mu})x + (M^{++}_\tau)^2  (1 -x)} \nonumber \\
&&
- ~\frac{m^2_{\mu} |y_\Delta^{e\mu}|^2}{2 \pi^2}
 \int_0^1 dx ~\frac{x^2(1-x)}{m^2_{\mu} x^2
  + ((M^{++}_\tau)^2 - m^2_{\mu})x + m^2_{\tau} (1 -x)}. \label{amu_H3pp}  
\eea
\eesub

The absence of a chirality-flipping term in the contributions from $H_{\m,\tau}^{++}$ is expected and, therefore, one observes 
$\Delta a^{\Delta^{++}}_{\mu}< 0$. 

LFV decays of $\tau \rightarrow \bar{\mu} e e$ and $\tau \rightarrow \bar{e} \mu \mu$ are allowed by the underlying $\mathbb{Z}_3$ symmetry. Of these, the branching fraction formula for the former process is the same as Eq.~\eqref{BR_taumuee_A} in Scenario A.
The branching fraction for the latter is given by
\besub
\bea
{\rm BR}_{\tau \rightarrow \bar{e} \mu \mu} &=& 
\frac{|y_{\Delta}^{e \tau}|^2 |y_{\Delta}^{\mu \mu}|^2}{4 G_F^2 
(M_\tau^{++})^4} \label{tau_emumu}
~.
\eea
\eesub

The independent parameters here are $M^+_\a, M^{++}_i, v_{\a}^{}, y_S^{\mu \tau}$, $y_S^{e e}$ and 
$\theta$. The muon $g-2$ is most sensitive to $M_{i}^{++}, y_{\Delta}^{\mu \tau}$ and $y_S^{\mu \tau}$. Among these, 
$y_{\Delta}^{ee}$ and $y_{\Delta}^{\mu \tau}$ can be fixed by the neutrino mass matrix elements as $y_{\Delta}^{ee} = \frac{m^{ee}_\nu}{\sqrt 2 v_e}$ and $y_{\Delta}^{\mu \tau} = \frac{m^{\mu \tau}_\nu}{\sqrt 2 v_e}$.
The following model parameter variation is made:
\besub
\bea
&&
500 ~\text{GeV} < M_1^{++} < 5 ~\text{TeV} ~, \\
&&
|y_S^{\mu \tau}|, |y_S^{e e}| < \sqrt{4 \pi} ~,\\
&&
10^{-14} ~\text{GeV} < v_{e}^{} < 10^{-4} ~\text{GeV} ~.
\eea
\eesub

In an approach similar to Scenario A, the representative values $\Delta M$ = 50~GeV, 100~GeV, $M_1^+ = M_1^{++}$ and $\theta = \frac{\pi}{4}, \frac{\pi}{10}$ are assigned. The remaining model parameters contribute only at subleading order to $\Delta a_\mu$, leading us to fix $M_\mu^+ = 1$~TeV, $M_\tau^+ = 1.2$~TeV, and $M_\mu^{++} = M_\tau^{++} = 1.1$~TeV. 
The neutrino oscillation parameters are fixed to their central values as shown in Eq.~(\ref{nuparam_fixed}).

With Eq.~(\ref{tau_emumu}), ${\rm BR}_{\tau \rightarrow \bar{e} \mu \mu} < 10^{-8}$ is translated to
\bea
v_\mu \gtrsim \frac{v}{M_\tau^{++}} 
\sqrt{ \frac{|m_\nu^{e \tau}|}{\text{1 eV}} \frac{|m_\nu^{\mu \mu}|}{\text{1 eV}}} \times 10^{-7} ~\text{GeV}
~.
\eea 
For typical values of $M_\tau^{++} \simeq 1$~TeV and $|m_{\nu}^{e \tau}|, ~|m_{\nu}^{e \tau}| \simeq 0.1$~eV, we get $v_\mu \gtrsim 2.5 \times 10^{-9}$~GeV. We have therefore chosen $v_\mu = v_\tau = 10^{-8}$~GeV in this analysis to ensure a suppressed rate for $\tau \rightarrow \bar{e} \mu \mu$. Also, once all the triplet VEV's are fixed, all $y_\Delta^{\a \b}$ can be determined from the neutrino mass matrix. Note that this choice for $v_{e}^{}$ and $v_{\tau}^{}$ renders the contributions of $H_\mu^{++}$ and $H_\tau^{++}$ to $\Delta a_\mu$ negligible. In the following, we plot the parameter points favoring a $\Delta a_\mu$ in the 2$\sigma$ interval, a perturbative $\l_7$ and sufficiently small decay rate in the $\tau \rightarrow \bar{\mu} e e$ channel in various planes of the parameter space.

It is important to highlight how the present scenario numerically differs from Scenario A. First, the allowed parameter space in the current scenario shows similar trends as in the case of Scenario A (see Fig.~\ref{f:ydelmutau-mH1pp}), much due to a common mechanism to explain $\Delta a_\mu$. However, a main difference lies in the fact that $y_{\Delta}^{\mu \tau}$ is now proportional to $m_{\nu}^{\mu \tau}$. This correlation gives the restriction $|\text{Re}(y_{\Delta}^{\mu \tau})| < 0.8$ for $\Delta M = 100$ GeV. 
On the other hand, the corresponding bound is more relaxed in case of Scenario A, as seen by a comparison between Fig.~\ref{f:m1pp-ydelmutau_B} and Fig.~\ref{f:ydelmutau-mH1pp}. In a way, Fig.~\ref{f:m1pp-ydelmutau_B} can be seen as a constrained version of Fig.~\ref{f:ydelmutau-mH1pp}.
Given that the chirality flip contribution is proportional to $\sim \Delta M s_{\theta} c_{\theta} y_{\Delta}^{\mu \tau} y_S^{\mu \tau}$, a lower $|y_{\Delta}^{\mu \tau}|$ in Scenario B calls for a higher $\Delta M$ and/or a lower $M_1^{++}$ in order to maintain the muon enhancement at the same magnitude.

\begin{figure}[tbhp]
\begin{center}
%
\includegraphics[scale=0.50]{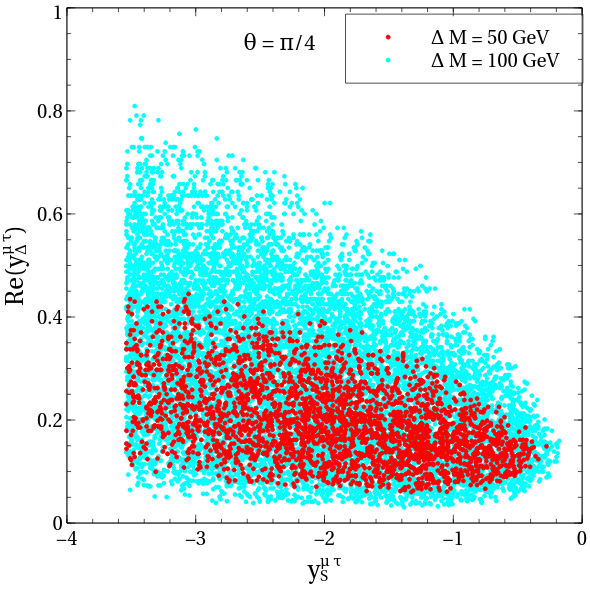}~~~ 
\includegraphics[scale=0.50]{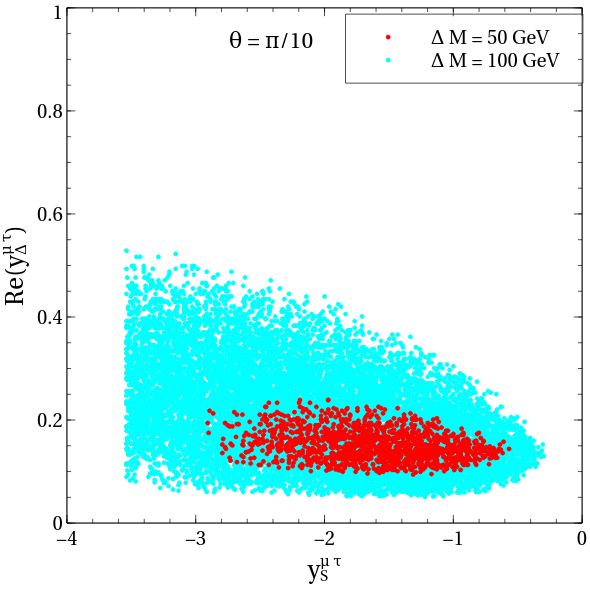}~~~\\ 
\caption{The allowed parameter space maintaining $\Delta a_\mu$ within its 2$\sigma$ range and ${\rm BR}_{\tau \rightarrow \bar{\mu} e e}$ within the quoted limit in the $y_S^{\mu \tau} - \text{Re}(y_\Delta^{\mu \tau})$ plane for $\theta = \frac{\pi}{4}$ (left) and $\frac{\pi}{10}$ (right). A normal neutrino mass hierarchy is assumed. The color coding is explained in the legends.}
\label{f:ySmutau-ydelmutau_B}
\end{center}
\end{figure}

\begin{figure}[tbhp]
\begin{center}
\includegraphics[scale=0.50]{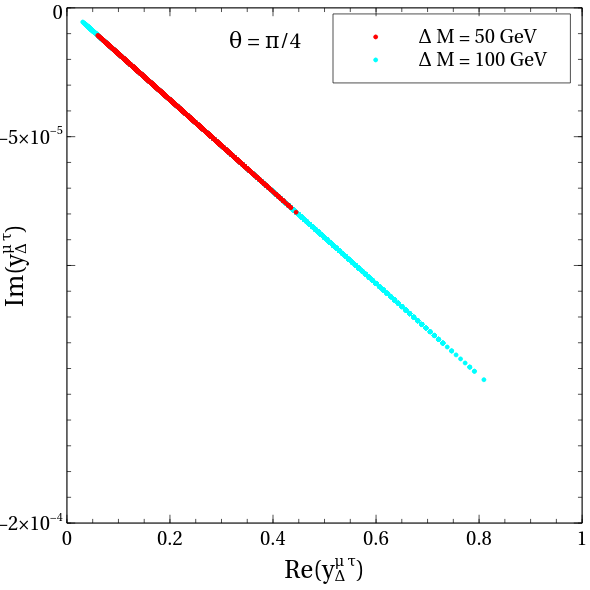}~~~ 
\includegraphics[scale=0.50]{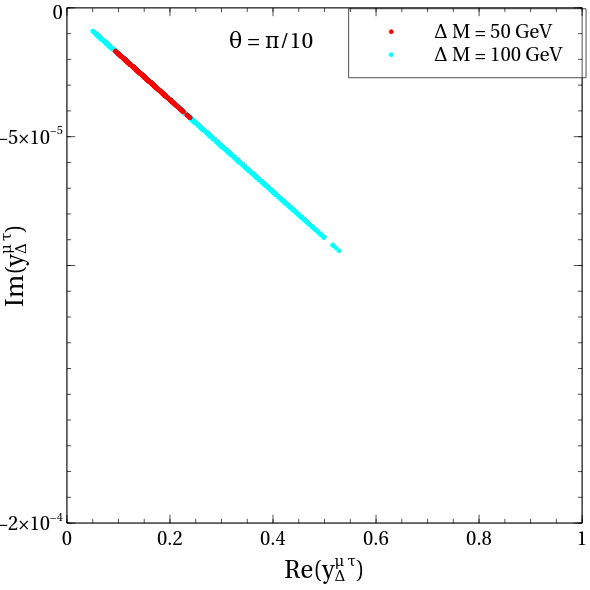}~~~\\
\caption{Scatter points maintaining $\Delta a_\mu$ within its 2$\sigma$ range and $\text{BR}_{\tau \rightarrow \bar{\mu} e e}$ within the quoted limit plotted in the $\text{Re}(y_\Delta^{\mu \tau}) - \text{Im}(y_\Delta^{\mu \tau})$ plane for $\theta = \frac{\pi}{4}$ (left) and $\frac{\pi}{10}$ (right). A normal neutrino mass hierarchy is assumed. The color coding is explained in the legends.}
\label{f:Rydelmutau-Iydelmutau_B}
\end{center}
\end{figure}

 
In Fig.~\ref{f:ySmutau-ydelmutau_B} together with Fig.~\ref{f:Rydelmutau-Iydelmutau_B}, 
we show the allowed parameter space in the $y_{\Delta}^{\mu\tau}$--$y_{S}^{\mu\tau}$ plane. 
These plots characterize the contributions to muon $g-2$, 
and should be compared with Fig.~\ref{f:ydelmutau-ySmutau} in Scenario A. 
Because of the proportionality relation $m_{\nu}^{\mu\tau} \propto y_{\Delta}^{\mu\tau}$, the parameter
$y_{\Delta}^{\mu\tau}$ is constrained more severely in Scenario B for a given $v_{e}^{}$.
In order to fit the neutrino oscillation data, the real part and the imaginary part of $y_{\Delta}^{\mu\tau}$ are strongly correlated, as shown in Fig.~\ref{f:Rydelmutau-Iydelmutau_B}. 
This correlation is somewhat similar to the relation between $\mu_{\a\b}^{}$ and $m_{\nu}^{\a\b}$ 
in Scenario A (see also Figs.~\ref{f:NH} and \ref{f:IH}). 
%

\begin{figure}[tbhp]
\begin{center}
\includegraphics[scale=0.50]{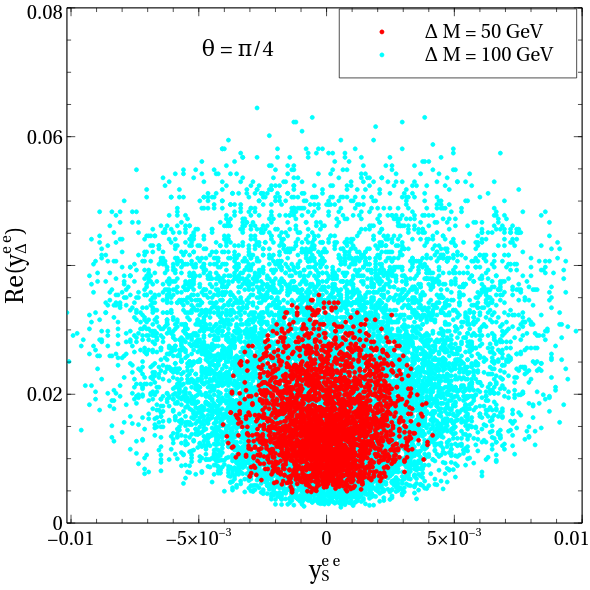}~~~ 
\includegraphics[scale=0.50]{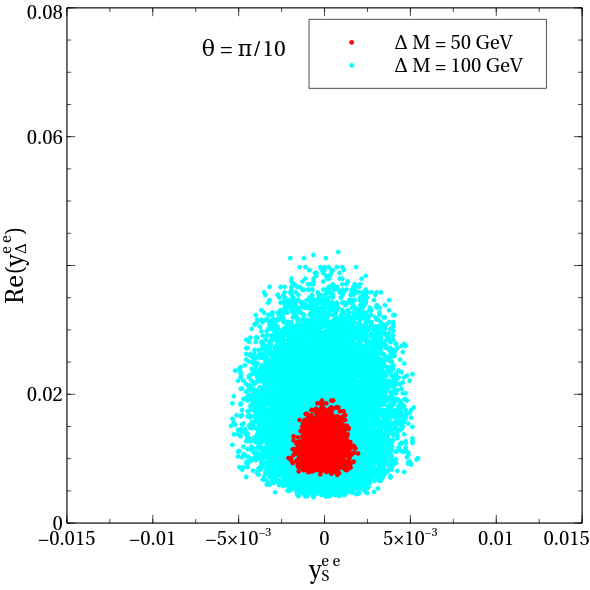}~~~\\
\caption{Scatter points maintaining $\Delta a_\mu$ 
within its 2$\sigma$ range and ${\rm BR}_{\tau \rightarrow \bar{\mu} e e}$ within the quoted limit plotted in the $y_S^{e e} - \text{Re}(y_\Delta^{e e})$
plane for $\theta = \frac{\pi}{4}$ (left) and $\frac{\pi}{10}$ (right). A normal neutrino mass hierarchy is assumed. The color coding is explained in the legends.}
\label{f:ySee-ydelee_B}
\end{center}
\end{figure}

\begin{figure}[tbhp]
\begin{center}
\includegraphics[scale=0.50]{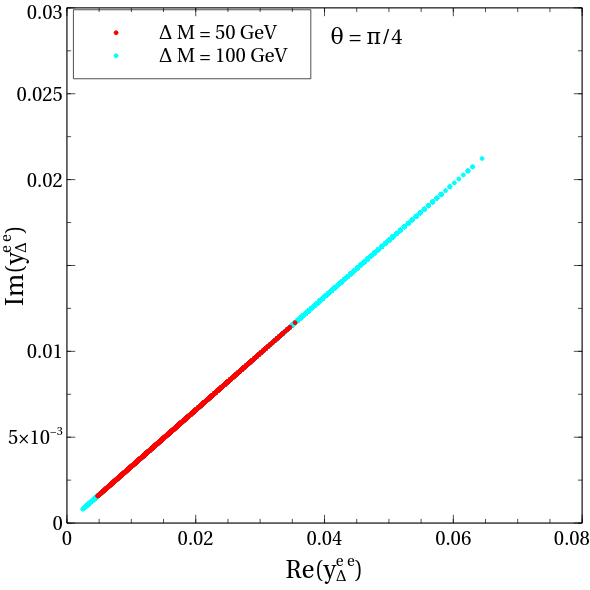}~~~ 
\includegraphics[scale=0.50]{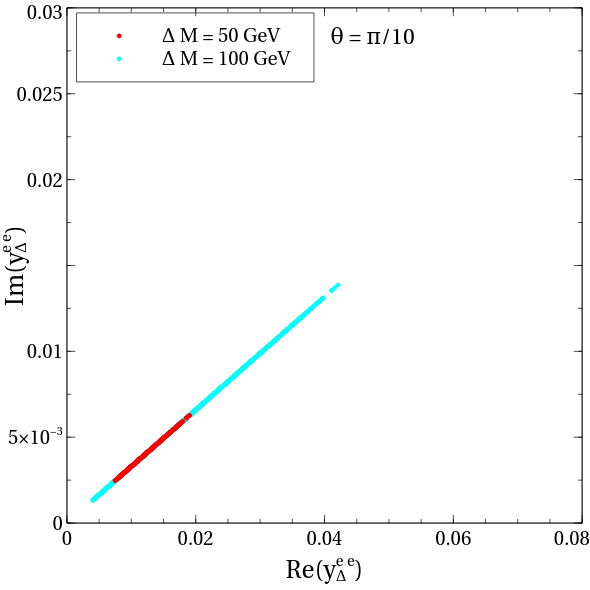}~~~\\  
\caption{Points maintaining $\Delta a_\mu$ 
within its 2$\sigma$ range and $\text{BR}_{\tau \rightarrow \bar{\mu} e e}$ within the quoted limit plotted in the $\text{Re}(y_\Delta^{e e}) - \text{Im}(y_\Delta^{e e})$
plane for $\theta = \frac{\pi}{4}$ (left) and $\frac{\pi}{10}$ (right). A normal neutrino mass hierarchy is assumed. The color coding is explained in the legends.}
\label{f:Rydelee-Iydelee_B}
\end{center}
\end{figure}

The couplings $y_\Delta^{\mu \tau}$ and $y_\Delta^{e e}$ that enter the expression for the $\tau \rightarrow \bar{\mu} e e$ branching fraction are dictated by the size of the $(ee)$ and $(\mu \tau)$ neutrino mass matrix elements, respectively. Therefore, the choice of the neutrino mass hierarchy becomes crucial in the analysis. In the case of NH, $|m_\nu^{ee}| \sim \mathcal{O}(10^{-3})$ eV throughout the entire space allowed by the oscillation data. However, the same is $\mathcal{O}(10^{-2})$ eV for the IH case, causing the $\tau \rightarrow \bar{\mu} e e$ branching ratio to overshoot the allowed limit by a factor of $\sim \mathcal{O}(10^2)$.  Consequently, no parameter point survives in the case of IH when the muon $g-2$ and LFV constraints are considered simultaneously. 


We read from Fig.~\ref{f:ySee-ydelee_B} that the bound on $|y_{\Delta}^{ee}|$ is about 0.07 for $\theta = \frac{\pi}{4}$ and settles to about 0.05 for $\theta = \frac{\pi}{10}$.  These numbers are close to the corresponding numbers in Scenario A. However, $y_S^{ee}$ is more constrained in the present case.
This is attributed to the fact that $|y_{\Delta}^{\mu \tau}|$ is more tightly constrained in Scenario B. 
${\rm BR}_{\tau \rightarrow \bar{\mu} e e}$ therefore allows the bound on $y_\Delta^{ee}$ to be loosened accordingly. For completeness, we also display the imaginary part of $y_\Delta^{e e}$ in Fig.~\ref{f:Rydelee-Iydelee_B}.

\begin{figure}[t]
\begin{center}
\includegraphics[scale=0.50]{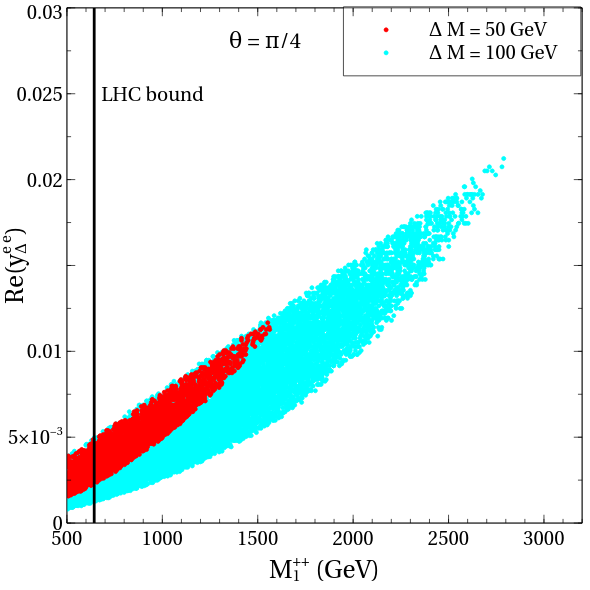}~~~ 
\includegraphics[scale=0.50]{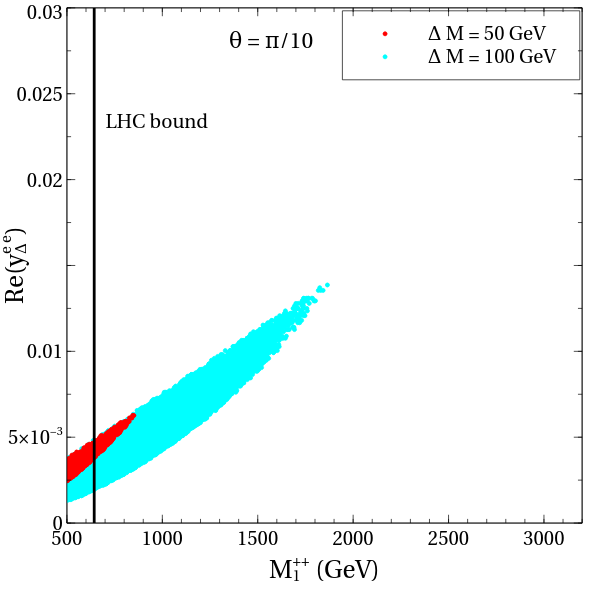}~~~\\ 
\caption{Scatter points maintaining $\Delta a_\mu$ within its 2$\sigma$ range and ${\rm BR}_{\tau \rightarrow \bar{\mu} e e}$ within the quoted limit plotted in the $M_1^{++} - \text{Re}
(y_\Delta^{e e})$
plane for $\theta = \frac{\pi}{4}$ (left) and $\frac{\pi}{10}$ (right). A normal neutrino mass hierarchy is assumed. The color coding is explained in the legends. The region left to the black line is disallowed by dilepton searches at the LHC.}
\label{f:m1pp-ydelee_B}
\end{center}
\end{figure} 

In Fig.~\ref{f:m1pp-ydelee_B}, we show the allowed parameter space in the $y_{\Delta}^{ee}$--$M_1^{++}$ plane.  Here, we take $\theta = \frac{\pi}{4}$ and $\Delta M = 50$~GeV, and find that $\Delta a_\mu$ in its 2$\sigma$ range disfavors $M_1^{++} \gtrsim 1.6$~TeV. The corresponding disfavored range stands at $M_1^{++} \gtrsim 3.7$~TeV in Scenario A. In the same logic, $\Delta M = 10$~GeV is disfavored in Scenario B as it does not provide the required $\Delta a_\mu$ enhancement.  A reduction in the parameter space after switching from the maximal mixing ($\theta = \frac{\pi}{4}$) to another angle ($\theta = \frac{\pi}{10}$ here) is expected and seen in all the plots. 

\begin{figure}[t]
\begin{center}
%
\includegraphics[scale=0.50]{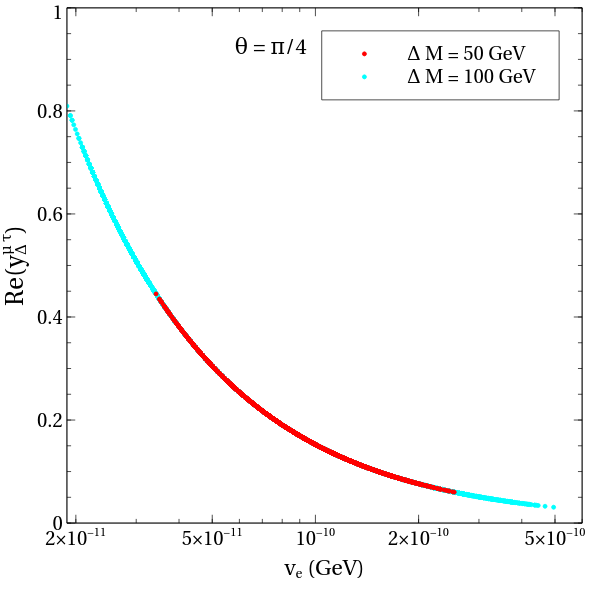}~~~ 
\includegraphics[scale=0.50]{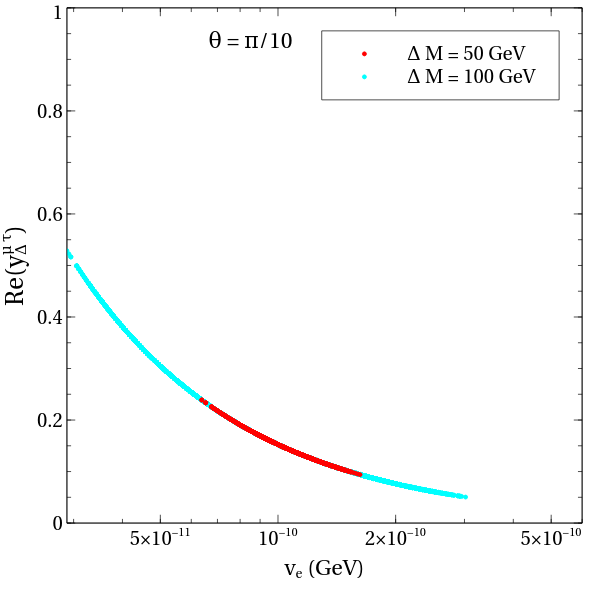}~~~\\ 
\caption{Scatter points maintaining $\Delta a_\mu$ 
within its 2$\sigma$ range and $\text{BR}_{\tau \rightarrow \bar{\mu} e e}$ within the quoted limit in the $v_{e} - \text{Re}(y_\Delta^{\mu \tau})$
plane for $\theta = \frac{\pi}{4}$ (left) and $\frac{\pi}{10}$ (right). A normal neutrino mass hierarchy is assumed. The color coding is explained in the legends.}
\label{f:ve-ydelmutau_B}
\end{center}
\end{figure}

The triplet VEV $v_{e}^{}$ turns out to be bounded from both above and below in Scenario B, as seen in Fig.~\ref{f:ve-ydelmutau_B}. This is because the maximally (minimally) allowed values of $y_\Delta^{\mu \tau}$ and $y_\Delta^{e e}$ passing the constraints come from the minimum (maximum) of $v_{e}^{}$ for given $m_\nu^{\mu \tau}$ and $m_\nu^{e e}$. Again, this is in contrast with Scenario A where there is no such bound.

As a closing remark, Scenario B is more constrained than Scenario A, in spite of having a larger number of scalar degrees of freedom. This is because of the Type-II-like $y_\Delta \sim \frac{m_\nu}{v_\Delta}$ relation in the scenario. Therefore, the sizes of the $(ee)$ and $(\mu\tau)$ elements in the neutrino mass matrix are crucial in shaping up the allowed parameter space. Going from Scenario A to Scenario B, 
the IH becomes disallowed. And this is found to hold true even if the 
neutrino oscillation parameters are varied within their allowed ranges. However, the parameter regions corresponding to NH open up a bit further in that case.
Scenario A enjoys more freedom precisely due to the presence of
$\mathbb{Z}_3$-breaking trilinear parameters. An appropriate choice of these parameters can reproduce both NH as well as IH without conflict with the muon $g-2$ anomaly and LFV decay bounds.

\section{Summary and conclusions}\label{summary}

The main theme of the present work is an explanation of the muon $g-2$ anomaly by arranging for a mixing between the doubly charged scalar belonging to an $SU(2)_L$ triplet and a doubly charged $SU(2)_L$ scalar singlet. The doubly charged mass eigenstates then couple to both chiralities of leptons. In such a case, the chirality flip in the muon $g-2$ loops can induce the requisite positive contribution so as to accommodate the anomaly. We have proposed two models (Scenario A and Scenario B) to investigate this effect. We have also sought to address non-zero neutrino mass and to satisfy the LFV decay constraints at the same time.

In Scenario A, the SM scalar sector is augmented by a complex scalar triplet $\Delta$, a doubly charged scalar singlet $k^{++}$, and three singly charged scalar singlets $k^+_e,k^+_\mu,k^+_\tau$. A softly broken $\mathbb{Z}_3$
symmetry is imposed under which $k^+_e,k^+_\mu,k^+_\tau$ have charges $1,\omega,\omega^2$, respectively, while $\Delta, k^{++}$ have charge 1.  However, soft $\mathbb{Z}_3$-breaking quadratic and trilinear terms are allowed, thereby causing the singly charged scalars to mix. Neutrino mass arises at the two-loop level and, therefore, this framework is a generalization of the well-known Zee-Babu model. The main findings in this scenario are the following:

\begin{itemize}

\item Owing to the $\Delta$--$k^{++}$ mixing, the dipole term corresponding to muon $g-2$ receives a boosted contribution. More precisely, this is due to the chirality flip and a logarithmic term in the loop amplitudes. It therefore becomes possible to address the muon $g-2$ anomaly in this framework.

\item The singlet-triplet scalar mixing plays a pivotal role also in the case of neutrino mass. A non-zero mixing induces a new two-loop amplitude that enjoys a chirality enhancement over the usual Zee-Babu-like diagram. In this paper, we have calculated the two-loop integrals exactly, including one which to our knowledge has not been done before. We have shown that by a suitable choice of the soft $\mathbb{Z}_3$ breaking trilinear parameters, it is possible to satisfy the present neutrino oscillation data. We have demonstrated it through benchmark points that agree with normal and inverted mass hierarchies. 

\item In the absence of $\mathbb{Z}_3$-breaking quadratic terms, the only non-trivial LFV process is $\tau \rightarrow \bar{\mu} e e$. We have shown that the rate of this process can be maintained within the allowed limit in the parameter region that accounts for the muon $g-2$ anomaly.  

\item The triplet VEV is allowed to take a wide range of values.

\end{itemize}

In Scenario B, three scalar triplets $\Delta_e, \Delta_\mu, \Delta_\tau$ having $\mathbb{Z}_3$ charges 1, $\omega,~\omega^2$ respectively and one doubly charged scalar singlet $k^{++}$, each having $\mathbb{Z}_3$ charge 1, are introduced. 
A violation of $\mathbb{Z}_3$ through soft terms is necessary here as neutrino mass is generated at the tree level when the triplets acquire VEV's. Once again, mixing between between the doubly charged state of $\Delta_e$ and $k^{++}$ occurs after EWSB. Some salient features of the allowed parameter region in this case are as follows:
\begin{itemize}

\item In this case, the $\Delta_e$--$k^{++}$ mixing also paves the way for a chirality flipping contribution in the muon $g-2$ loops. The requisite enhancement in muon $g-2$ is therefore generated in a manner similar to the previous scenario.

\item In the case of a normal neutrino mass hierarchy, the parameter space favoring an enhanced muon $g-2$ also complies with the bounds on the branching fractions of $\tau \rightarrow \bar{\mu} e e$ and $\tau \rightarrow \bar{e} \mu \mu$, the only non-vanishing LFV modes in this scenario.

\item The present scenario disfavors an inverted neutrino mass hierarchy. This is attributed to the fact that the $m_\nu^{e e}$ value associated with the IH is typically larger than the corresponding NH value by at least an order of magnitude. As a result, the rate of $\tau \rightarrow \bar{\mu} e e$ is often predicted above the permitted limit.

\item Unlike in Scenario A, the triplet VEV gets bounded from both ends in the process of reconciling the muon $g-2$ anomaly with LFV constraints.

\end{itemize}

The introduction of any dimension-$2$ $\mathbb{Z}_3$-breaking terms in such scenarios will lead to quadratic mixing between the scalars and, therefore, turn on the loop-induced $l_\a \rightarrow l_\b \gamma$ LFV processes. For both Scenario A and Scenario B, singly charged and doubly charged scalars will be running in the loops. However, the $\mathbb{Z}_3$-violating Yukawa interactions so induced will obviously be proportional to the magnitude of the quadratic mixing. Hence, such LFV rates can be easily controlled by keeping the magnitude of the $\mathbb{Z}_3$-breaking terms sufficiently small.

Finally, a remark on possible collider signatures of these models is in order. The strengths of the $\mu \tau$ Yukawa couplings of the doubly charged scalars in both scenarios are found to be much larger than the corresponding $e e$ strength. In such a case, $pp \to H^{++}_1 H^{--}_2$ followed by $H^{++}_{1,2} \to$ $\mu^+ \tau^+$ can give rise to a pair of like-sign dilepton $\mu \tau$ with an invariant mass peaking around $M^{++}_1$ and $M^{++}_2$, respectively. For a sizeable mass gap, these invariant mass peaks would share no overlap. A resolution of these two peaks can enable one to distinguish the proposed scenarios from the pure Type-II model.

\acknowledgments

This research of CWC was supported
by the Ministry of Science and Technology of Taiwan under Grant No. MOST 104-2628-M-002-014-MY4. The work of KT is supported by JSPS Grant-in-Aid for Young Scientists (B) (Grant No. 16K17697) 
and the MEXT Grant-in-Aid for Scientific Research on Innovation Areas (Grant No. 18H05543). 
NC thanks Titas Chanda for an important computational help.

\appendix
\section{Appendix}

This section contains various analytical expressions related to $\Delta a_\mu$ and neutrino mass.

\subsection{Muon $g-2$ integrals}
The $H_i^{++}$ contributions to $\Delta a_\mu$ contain the following integrals with $M_S$ denoting $M_{1,2}^{++}$:

\begin{align}
&\int_0^1 dx ~\frac{x^2-x^3}{m^2_{\mu} x^2 + (m^2_{\tau} - m^2_{\mu})x + M_S^2  (1 -x)} \nonumber \\
&=  \Big[2 M_S^6 + 3 M_S^4 m_\tau^2 - 6 M_S^2 m_\tau^4 + 
m_\tau^6 - 6 M_S^4 m_\tau^2 ~\text{log} \frac{m^2_\tau}{M_S^2}\Big]/
\Big[6 (M_S^2 - m_\tau^2)^2\Big] ~, \\ 
&\int_0^1 dx ~\frac{x^2}{m^2_{\mu} x^2 + (m^2_{\tau} - m^2_{\mu})x + M_S^2  (1 -x)} \nonumber \\
&= \Big[(M_S^2 - m_\tau^2) \Big( -3 M_S^4 + 4 M_S^2 m_\tau^2
 - m_\tau^4 + 2 M_S^4 ~\text{log} \frac{m^2_\tau}{M_S^2}\Big)\Big]/
\Big[2(M_S^2 - m_\tau^2)^2\Big] ~, \\ 
&\int_0^1 dx ~\frac{x^2(x-1)}{m^2_{\mu} x^2
  + (M_S^2 - m^2_{\mu})x + m^2_{\tau} (1 -x)} \nonumber \\
&= -\Big[M_S^6 - 6 M_S^4 m_\tau^2 + 7 M_S^2 m_\tau^4 - 
2 m_\tau^6 + 6 M_S^4 m_\tau^2 ~\text{log} \frac{m^2_\tau}{M_S^2}\Big]/
\Big[6 (M_S^2 - m_\tau^2)^2\Big] ~, \\ 
&\int_0^1 dx ~\frac{x(x-1)}{m^2_{\mu} x^2
  + (M_S^2 - m^2_{\mu})x + m^2_{\tau} (1 -x)} \nonumber \\
&= -\Big[(M_S^2 - m_\tau^2) \Big(M_S^4
  m_\tau^4 - 2 M_S^2 m^2_\tau ~\text{log} \frac{m^2_\tau}{M_S^2}\Big)\Big]/
\Big[2 (M_S^2 - m_\tau^2)^2\Big] ~.
\end{align}

\subsection{Evaluation of $I_k(m_1,m_2,m,m_c,m_d)$}

We use the notation in Ref.~\cite{McDonald:2003zj} when calculating the two-loop integrals connected to neutrino mass generation:  
\besub
\bea
(m_1|m_2|m) &=& \int d^d p_E  d^d q_E~
 \frac{1}{(p_E^2 + m^2_1)(q_E^2 + m^2_2)((p_E + q_E)^2 + m^2)} ~, \\ 
(2 m_1|m_2|m) &=& \int d^d p_E  d^d q_E~
\frac{1}{(p_E^2 + m^2_1)^2(q_E^2 + m^2_2)((p_E + q_E)^2 + m^2)} \\ 
&=& -\pi^4 \Big[-\frac{2}{\epsilon^2} + \frac{1}{\epsilon}
(1 - 2 \gamma_E - 2 \text{log}(\pi m^2_1))\Big] \nonumber \\
&&
-\pi^4 \Big[-\frac{1}{2} - \frac{\pi^2}{12} - \gamma_E^2
 + (1 - 2 \gamma_E) \text{log}(\pi m^2_1) - \text{log}^2(\pi m^2_1) - f(m_1,m_2,m)\Big] \nonumber \\
&& 
  + \mathcal{O}(\epsilon) ~,
\eea
\eesub
where
\besub
\bea
f(m_1,m_2,m_3) &=& \int_0^1 dx \Big( \text{Li}_2(1 - \mu^2) - \frac{\mu^2 \text{log} \mu^2}{1 - \mu^2}\Big) ~, \\
\text{and }~
\mu^2 &=& \frac{m_2^2 x + m^2 (1 - x)}{x(1-x) m_1^2}
\eea
\eesub

The contribution of $k_{++}$ to neutrino mass is given by
\begin{align}
&I_k(m_1,m_2,m,m_c,m_d) \nonumber \\
&= \int \frac{d^d p_E}{(2\pi)^d} \frac{d^d q_E}{(2\pi)^d}~ 
 \frac{m_c m_d}{(p_E^2 + m^2_1)(p_E^2 + m^2_c)(q_E^2 + m^2_2)(q_E^2 + m^2_d)((p_E + q_E)^2 + m^2)} \nonumber \\
&= \frac{1}{(2\pi)^8}\frac{m_c m_d}{(m_1^2 - m_c^2)(m_2^2 - m_d^2)} 
 \Big[(m_1|m_2|m) - (m_c|m_2|m) - (m_1|m_d|m)
  + (m_c|m_d|m)\Big] 
\end{align}

In $d = 4$ dimensions, the following holds
\bea
(m_1|m_2|m) &=& -\Big[m_1^2 (2 m_1|m_2|m)
 + m_2^2 (2 m_2|m_1|m) + m^2 (2 m|m_1|m_2)\Big]
\eea

Therefore, 
\besub
\bea
I_k(m_1,m_2,m,m_c,m_d) &=& \frac{1}{(2\pi)^8}\frac{1}{(3 - d)}
\frac{m_c m_d}{(m_1^2 - m_c^2)(m_2^2 - m_d^2)}   \nonumber \\
&&
\Big[m_1^2 (2 m_1|m_2|m) + m_2^2 (2 m_2|m_1|m) + m^2 (2 m|m_1|m_2) \nonumber \\
&&
-m_c^2 (2 m_c|m_2|m) - m_2^2 (2 m_2|m_c|m) - m^2(2 m|m_c|m_2) \nonumber \\
&&
-m_1^2(2 m_1|m_d|m) - m_d^2(2 m_d|m_1|m) - m^2(2 m|m_1|m_d) \nonumber \\
&&
+ m_c^2(2 m_c|m_d|m) + m_d^2(2 m_d|m_c|m) + m^2(2 m|m_c|m_d)
\Big] \\ \nonumber \\
&=& \frac{1}{(4\pi)^4}
\frac{-m_c m_d}{(m_1^2 - m_c^2)(m_2^2 - m_d^2)} \nonumber \\
&&
\Big[m_1^2 f(m_1,m_2,m) + m_2^2 f(m_2,m_1,m) + m^2 f(m,m_1,m_2) \nonumber \\
&&
-m_c^2 f(m_c,m_2,m) - m_2^2 f(m_2,m_c,m) - m^2 f(m,m_c,m_2) \nonumber \\
&&
-m_1^2 f(m_1,m_d,m) - m_d^2 f(m_d,m_1,m) - m^2 f(m,m_1,m_d) \nonumber \\
&&
+ m_c^2 f(m_c,m_d,m) \nonumber \\
&&
 + m_d^2 f(m_d,m_c,m) + m^2 f(m,m_c,m_d)\Big]
\eea
\eesub

Therefore, $I_k(m_1,m_2,m,m_c,m_d)$ is UV finite.

\subsection{Evaluation of $I_{\Delta}(m_1,m_2,m,m_c,m_d)$}

The contribution coming from $\delta_{++}$ involves the following integral:
%
\begin{align}
&I_{\Delta}(m_1,m_2,m,m_c,m_d) \nonumber \\
&=
-\int \frac{d^d p_E}{(2\pi)^d} \frac{d^d q_E}{(2\pi)^d}
 \frac{p_E.q_E}{(p_E^2 + m^2_1)(p_E^2 + m^2_c)(q_E^2 + m^2_2)(q_E^2 + m^2_d)((p_E + q_E)^2 + m^2)}
\end{align}


%

We define
\besub
\bea
D_1 &=& p_E^2 + m^2_1 \\
D_2 &=& q_E^2 + m^2_2 \\
D_c &=& p_E^2 + m^2_c \\
D_d &=& q_E^2 + m^2_d \\
D &=& (p_E + q_E)^2 + m^2
\eea
\eesub
and
\besub
\bea
&& I_{\Delta}(m_1,m_2,m,m_c,m_d) 
\nonumber \\
&&=
-\frac{1}{2}\int \frac{d^d p_E}{(2\pi)^d} \frac{d^d q_E}{(2\pi)^d}~ 
%
\Bigg[\frac{(D - m^2 - D_1 + m^2_1 - D_2 + m^2_2)}{D_1 D_c D_2 D_d D} \Bigg]
\\
\nonumber \\
&&= 
-\frac{1}{2}\int \frac{d^d p_E}{(2\pi)^d} \frac{d^d q_E}{(2\pi)^d}~
\Bigg[\frac{1}{D_1 D_c D_2 D_d}
 - \frac{1}{D_c D_2 D_d D} \nonumber \\
&& \qquad\qquad\qquad\qquad\qquad\qquad
 - \frac{1}{D_1 D_c D_d D}
  + \frac{(m^2_1 + m^2_2 - m^2)}
 {D_1 D_c D_2 D_d D}\Bigg]
\eea
\eesub

We split the second, third and fourth terms using partial fractions as
\besub
\bea
I_{\Delta}(m_1,m_2,m,m_c,m_d) &=& -\frac{1}{2}\int \frac{d^d p_E}{(2\pi)^d} \frac{d^d q_E}{(2\pi)^d}~ \Bigg[\frac{1}{D_1 D_c D_2 D_d}\Bigg] \nonumber \\
&& 
- \frac{1}{2}\frac{1}{(2\pi)^8} \frac{1}{(m_2^2 - m_d^2)}
\Big[(m_c|m_2|m) - (m_c|m_d|m)\Big] \nonumber \\
&&
- \frac{1}{2}\frac{1}{(2\pi)^8} \frac{1}{(m_1^2 - m_c^2)}
\Big[(m_1|m_d|m) - (m_c|m_d|m)\Big]\nonumber  \\
&&
- \frac{1}{2}\frac{1}{(2\pi)^8}\frac{(m_1^2 + m_2^2 - m^2)}{(m_1^2 - m_c^2)(m_2^2 - m_d^2)}
\Big[(m_1|m_2|m) - (m_1|m_d|m)\nonumber \\
&&
 - (m_c|m_2|m) + (m_c|m_d|m)\Big] \\ \nonumber \\
&=& -\frac{1}{2}\int \frac{d^d p_E}{(2\pi)^d} \frac{d^d q_E}{(2\pi)^d}~ \Bigg[\frac{1}{D_1 D_c D_2 D_d}\Bigg] \nonumber \\
&&
- \frac{1}{2}\frac{1}{(2\pi)^8}
\frac{1}{(m_1^2 - m_c^2)(m_2^2 - m_d^2)} \Bigg[
(m_1^2 + m_2^2 - m^2) (m_1|m_2|m) \nonumber \\
&&
+ (m^2 - m^2_2 - m^2_c)(m_c|m_2|m)
+ (m^2 - m^2_1 - m^2_d)(m_1|m_d|m)
\nonumber \\
&&
+ (m_c^2 + m_d^2 - m^2) (m_c|m_d|m)
\Bigg] \\ \nonumber \\
&=& -\frac{1}{2}\int \frac{d^d p_E}{(2\pi)^d} \frac{d^d q_E}{(2\pi)^d}~ \Bigg[\frac{1}{D_1 D_c D_2 D_d}\Bigg]
- \frac{1}{2}\frac{1}{(3 - d)}\frac{1}{(2\pi)^8} \frac{1}{(m_1^2 - m_c^2)(m_2^2 - m_d^2)}
\nonumber \\
&&
\Bigg[(m_1^2 + m_2^2 - m^2) \Big(m_1^2(2 m_1|m_2|m) + m_2^2 (2 m_2|m_1|m) + m^2 (2 m|m_1|m_2)\Big) \nonumber \\
&&
+ (m^2 - m_2^2 - m_c^2)\Big(m_c^2(2 m_c|m_2|m) + m_2^2 (2 m_2|m_c|m) + 
m^2 (2 m|m_c|m_2)\Big) \nonumber \\
&&
+ (m^2 - m_1^2 - m_d^2)\Big(m^2_1(2 m_1|m_d|m) + m^2_d(2 m_d|m_1|m)
 + m^2(2 m|m_1|m_d)\Big) \nonumber \\
&&
+ (m_c^2 + m_d^2 - m^2) \nonumber \\
&&
\Big(m^2_c(2 m_c|m_d|m) + m^2_d(2 m_d|m_c|m)
 + m^2(2 m|m_c|m_d)\Big)\Bigg]
\eea
\eesub

Note that $I_{\Delta}(m_1,m_2,m,m_c,m_d)$ is not UV-finite. However,
the combination that enters the neutrino mass, $I_{\Delta}(m_1^+,m_2^+,M_1^{++},m_c,m_d)
 - I_{\Delta}(m_1^+,m_2^+,m_2^{++},m_c,m_d)$, is.

\bibliographystyle{JHEP}
\bibliography{ref_zeebabu}

\end{document}